\documentclass[usenatbib]{mn2e}
\usepackage{natbibmnfix, graphicx, times}
\usepackage{epstopdf}

\newcommand{\rev}{}

\title[Recoiling supermassive black holes]{Effects of gravitational-wave recoil on the dynamics\\
and growth of supermassive black holes}

\author[Blecha and Loeb]{Laura Blecha and Abraham Loeb \\ 
Harvard University, Department of Astronomy, 60 Garden St., Cambridge, MA 02138, USA\\ Email: lblecha@cfa.harvard.edu, aloeb@cfa.harvard.edu}

\begin{document}
\maketitle

\begin{abstract}
Simulations of binary black hole mergers indicate that asymmetrical gravitational wave (GW) emission can cause black holes to recoil at speeds up to thousands of km s$^{-1}$.  These GW recoil events can dramatically affect the coevolution of recoiling supermassive black holes (SMBHs) and their host galaxies.  However, theoretical studies of SMBH-galaxy evolution almost always assume a stationary central black hole.  In light of the numerical results on GW recoil velocities, we relax that assumption here and consider the consequences of recoil for SMBH evolution.  We follow the trajectories of SMBHs ejected in a smooth background potential that includes both a stellar bulge and a multi-component gaseous disk.  In addition, we calculate the accretion rate onto the SMBH as a function of time using a hybrid prescription of viscous ($\alpha$-disk) and Bondi accretion.  We find that recoil kicks between 100 km s$^{-1}$ and the escape speed cause SMBHs to wander {\rev through} the galaxy and halo for $\sim 10^6 - 10^9$ yr before settling back to the galactic center.  However, the mass accreted during this time is roughly constant at $\sim 10\%$ of the initial mass, independent of the recoil velocity.  This indicates that while large recoils may disrupt active galactic nuclei feedback processes, recoil itself is an effective means of regulating SMBH growth.  Recoiling SMBHs may be observable as spatially or kinematically offset quasars, but finding such systems could be challenging, because the largest offsets correspond to the shortest quasar lifetimes.  
\end{abstract}

\begin{keywords}
black hole physics -- gravitational waves -- accretion, accretion disks -- galaxies: kinematics and dynamics -- galaxies: evolution -- galaxies: active
\end{keywords}

\section{Introduction}
\label{sec:intro}

Supermassive black holes (SMBHs, $\ga 10^6$ M$_{\odot}$) have long been considered the probable central engines for quasars \citep{lynden69}.  Observations now indicate that they are ubiquitous in the centers of both active and quiescent galaxies (e.g. \citet{korric95, richst98, ferfor05}).  Considerable evidence links SMBHs closely to dynamical properties of their host galaxies, indicating that SMBHs and galaxies may have evolved concurrently.  Measured SMBH masses are closely correlated with the luminosity, and hence the mass, of their host galactic bulges \citep{korric95, magorr98, marhun03, merfer01}.  Also, a strong correlation has been observed between SMBH mass and stellar velocity dispersion in the bulge: $M_{\rm BH} \propto \sigma_*^4$ \citep{gebhar00,fermer00, tremai02}.  Recent surveys have found that a similar correlation may extend down to dwarf galaxies if their compact nuclei are interpreted as the analog of SMBHs in massive galaxies \citep{ferrar06, wehhar06, barth05}.  

These findings indicate that SMBHs offer a wealth of information about the formation and evolution of structure in our universe.  In this context, much theoretical work has been devoted to studying the details of SMBH accretion and how it relates to galaxy evolution \citep[e.g.,][]{silree98, wyiloe03, dimatt05, hopkin06}.  However, virtually all such studies have one common assumption:  that the black hole (BH) remains stationary at the center of the galaxy.  This is a reasonable assumption in light of the observed ubiquity of central compact objects in galaxies, although the observed sample contains only local galaxies, so in principle few constraints exist on the dynamics of SMBHs over the lifetimes of galaxies.  However, if two SMBHs coalesce after a major galaxy merger, gravitational-wave (GW) recoil can impart a large kick to the resultant merged BH \citep{peres62, bekens73, fitdet84}.  When the merging BHs have unequal masses or spins, asymmetrical emission of gravitational waves creates a net linear momentum flux at merger, causing the merged BH to recoil in the opposite direction.  This effect has long been known as an interesting relativistic phenomenon, but its importance as an astrophysical phenomenon remained very uncertain until recently; the best estimates from approximate methods indicated that kick velocities were most likely to be small, $v_{\rm k} \la {\rm few} \times 100$ km s$^{-1}$  \citep{blanch05, damgop06}.  Results from simulations using full numerical relativity have demonstrated that {\rev if high BH spins are included,} kicks up to 4000 km s$^{-1}$ are possible with certain mass and spin configurations  \citep{campan07a, campan07b}.  These velocities easily exceed the escape speed of any known galaxy, so the ramifications of large recoil kicks would clearly be dramatic.  Not only will the position of the BH in the galaxy vary, but also the rate of fueling and the amount and type of active galactic nucleus (AGN) feedback generated.  For example, according to the Bondi accretion model, which is commonly used in simulations, the accretion rate is inversely proportional to the velocity of the accretor cubed \citep{bonhoy44}.  

The velocity of recoil kicks depends strongly on the spin parameters, spin alignment, and mass ratio of the merging BHs.  The distributions of these quantities in the population of merging galaxies are highly uncertain.  \citet{bogdan07} have suggested that torques in a circumbinary gas disk may align the BH spins with the disk orbital plane, causing the recoil distribution to be skewed toward in-plane kicks (and consequently low kick velocities).  {\rev \citet{schbuo07}, \citet{campan07a}, \& \citet{baker08} have each calculated probability distributions} of recoil kick velocities assuming randomly distributed BH mass ratios and {\rev spin orientations}.  We allow for both {\rev aligned and random spins} by considering a variety of kick inclinations.  Observing signatures of GW recoil in the future would be one way to constrain these distributions.

 A common result of early SMBH merger simulations was the so-called ``final parsec problem", whereby inspiralling SMBHs ``stall"  at a characteristic radius of about a parsec after hollowing out the host galaxy's core and before gravitational wave (GW) emission becomes dominant \citep[e.g.,][]{milmer01, yu02}.  This has been referred to as a ``problem" because until very recently \citep[see][]{rodrig06}, no direct evidence for any such unmerged binary SMBHs existed, let alone a large population of them.  Numerous viable solutions to this problem have been introduced, for example by adding a slightly triaxial potential or a nonnegligible gas fraction \citep[e.g.,][]{gerbin85, yu02, berczi06, escala04, goumir97}.  It is certainly possible, however, that in ``dry" mergers of galaxies containing small gas fractions, the SMBH inspiral time would exceed the time for another major merger to occur.  This could result in formation of a triple SMBH system, which may impart a significant kick to one of the BHs \citep[e.g.,][]{hutree92, xuost94}.  {\rev Recently, the first fully relativistic simulations of close triple-BH encounters demonstrated that substantial kicks can indeed occur \citep{campan08}.}  Using non-relativistic simulations, \citet{hofloe07} conducted a {\rev statistical} analysis of the triple-SMBH problem; they found that most three-body encounters cause the binary BH to merge {\rev without ejecting the third BH, and} that ejection of the third BH {\rev (which may also be followed by the merger of the other two)} occurs in less than half of all encounters.  It should be noted that if triple-SMBH systems do form in galaxies and if the most likely outcome is an SMBH merger, then GW recoil kicks could be at least as likely to result from this configuration as dynamical kicks to the third BH.    

\citet{madqua04} and \citet{loeb07} have considered some of the observational signatures we might expect from these sources, such as spatially and kinematically offset quasars.  These could be either SMBHs that are ejected from their host galaxies and carry an accretion disk along, or wandering SMBHs on bound trajectories that may significantly disrupt galactic structure.  Because no known quasars have been conclusively identified as such, we can assume the fraction of offset quasars is relatively small.  Similarly, \citet{volont07} considered the limits placed on ejected BHs by the $M_{\rm BH} - \sigma_*$ relation.  She concluded that ejection of SMBHs could not have been a very common occurrence throughout most of the history of the universe, or we would observe considerably more scatter in BH-galaxy relations.  However, Volonteri points out that in large halos at high redshift, mergers were more frequent and recoil kicks, which are independent of absolute mass, could more easily have ejected BHs from the relatively small host galaxies.  

{\rev \citet{guamer08} studied GW recoil with a set of N-body simulations following the motion of a kicked SMBH in a stellar potential.  They find that BHs on bound orbits have long oscillation timescales, up to $\sim 1$ Gyr, including a phase of low-amplitude oscillations in the stellar core.  A large stellar core is scoured out by these oscillations, in some cases larger than the cores that binary SMBHs are expected to produce.  \citet{guamer08} also discuss the possible observational signatures of such oscillating BHs, such as displaced AGN and offset jets.}    
 
In a recent paper, \citet{korlov08} have focused on a different aspect of GW recoil, namely its effect on galaxy morphology.  Using GADGET-2, they {\rev also} find that BHs on bound orbits have long oscillation timescales and conclude that recoil events can produce significant asymmetry in galactic disks.  We note, however, that by definition, any galaxy containing a recoiling SMBH will have recently undergone a merger, which will also produce asymmetric morphology. 

\citet{magain05} discovered a quasar with no apparent host galaxy located near an Ultra-Luminous Infrared Galaxy (ULIRG), which had likely undergone a recent merger.  A number of papers discussed whether this might be an ejected SMBH \citep{haehne06, merrit06, hofloe06}.  This does not appear to be the case for this system, however.  Among other factors, the quasar has a significant narrow-line region indicating the presence of more gas than could be ejected with the SMBH, and recently, direct evidence for a host galaxy has been observed \citep{feain07}.  

While this paper was undergoing revision, \citet{komoss08} announced that the quasar SDSSJ0927+2943 is a candidate recoiling SMBH with a velocity of 2650 km s$^{-1}$.  They have observed an offset between the broad emission lines and {\rev some of the} narrow emission lines in this unusual quasar's spectrum, which they suggest can be explained if the broad line region is gravitationally bound to a recoiling BH and the narrow line region is left behind in the galactic disk.  2650 km s$^{-1}$ is close to the maximum possible recoil kick velocity, {\rev and the actual velocity may be even higher if there is a velocity component perpendicular to the line of sight.  Thus,} this proposed recoil event is certainly not a typical event.  
 
\citet{bonshi07} conducted a targeted search for GW recoil events.  They examined a quasar population for Doppler shifts in spectral lines that would indicate a moving source and found a null result.  However, this method applies only to high-velocity BHs that are active quasar sources.  Additionally, because a BH on a bound trajectory spends a large fraction of its time at turnaround, such objects are statistically more likely to be observed with low velocities.  

In this paper, we focus on the dynamical aspect of SMBH accretion in the context of large kicks from GW recoil.  We emphasize that the details of BH accretion occur on scales below the current resolution of state-of-the-art simulations of galaxy mergers, and that the physics of turbulent viscosity driving this accretion cannot be simulated easily.  Thus, a semi-analytic approach is an appropriate first step toward understanding accretion onto recoiling BHs.  We use a set of semi-analytical models to follow the trajectories of SMBHs ejected with varying kick velocities in the potential of a galaxy containing both gas and stellar components; we also estimate the accretion rate for the moving SMBHs.  Our model and choice of parameters are outlined in \S~\ref{sec:model}.  The results of these calculations for our fiducial model are given in \S~\ref{ssec:fidmod}, while \S~\ref{ssec:massmod} \& \ref{ssec:fgasmod} detail the results of varying the SMBH mass and galaxy gas fraction.  In \S~\ref{ssec:obssig}, we consider the implications of our results for observational signatures of GW recoil.  \S~\ref{ssec:q_sensitivity} contains comments on the sensitivity of our results to our choice of parameters.  Finally, in \S~\ref{sec:disc} we summarize and discuss these results in the broader context of SMBH-galaxy coevolution.

\section{Model}
\label{sec:model}

We construct a galaxy model consisting of two distinct components: a spherical stellar bulge and a gaseous disk.  {\rev Note that in reality, the stellar and gas distributions may be highly asymmetric following a major merger.  Choosing symmetric distributions is therefore a simplification necessary for our semi-analytical approach, although if the SMBH merger timescale is long, the galaxy may in fact be significantly relaxed when the recoil occurs.}  The dark matter (DM) halo is ignored for the purpose of following SMBH orbits, because the halo profile will be fairly flat on the scale of the baryonic galaxy.  Also, on large scales the halo is expected to be significantly triaxial \citep{jinsut02, bulloc01}, {\rev so even though the DM halo would decrease the SMBH orbital decay time, it would also increase the chance of scattering the BH onto a less-centrophilic orbit, such that the BH would wander in the halo instead of returning to the galaxy center.  The details of this scattering are highly dependent on the shape of the triaxial potential, so we ignore this effect and consider it a source of uncertainty in our high-velocity recoil runs.}

\subsection{Stellar bulge}
\label{ssec:stellar}
The stellar bulge is assumed to follow an $\eta$-model density profile \citep{tremai04} with $\eta = 2.5$.  We choose this profile that is slightly flatter than a \citet{hernqu90} profile ($\eta = 2$) under the assumption that a galaxy merger remnant will have a flattened core due to stellar ejections during SMBH binary inspiral.  The total stellar bulge mass is estimated from the SMBH mass using the relation of \citet{marhun03},
\begin{equation}
M_{*} = 4.06\times10^{10} {\rm M}_{\odot} \left ( {M_{\rm BH} \over 10^8  {\rm M}_{\odot} }\right )^{1.04}.
\end{equation}
The scale radius, $a$, for the $\eta$-model profile is related to the SMBH mass in terms of the half-mass radius, $r_{1/2}$, 
\begin{equation}
a = r_{1/2} (2^{1/\eta} - 1).  
\end{equation}
An approximate relation for the ratio of effective (half-light) radius to half-mass radius is given by \citet[Eq. (17)]{dehnen93}
\begin{eqnarray}
{ R_{\rm e} \over r_{1/2} } \approx 0.7549 - 0.00439(3-\eta) + 0.00322(3-\eta)^2 \nonumber \\
 - 0.00182 (3-\eta)^3 \pm 0.0007.
\end{eqnarray}
The effective radius is $R_{\rm e} = G M_* / 3 \sigma_*^2$, and the one-dimensional velocity dispersion, $\sigma_*$, is calculated from the $M_{\rm BH} - \sigma_*$ relation \citep{tremai02}, 
\begin{equation}
\sigma_* = 220 \, {\rm km \, s^{-1}} \left ( { M_{\rm BH}\, \over \, 2\times10^8 {\rm M}_{\odot} } \right )^{0.249}.
\end{equation}
Thus, the scaling of the stellar model radii with BH mass is ($R_{\rm e}$, $r_{1/2}$, $a$) $\propto$ $M_{\rm BH}^{1/2}$.

Stellar dynamical friction on the ejected SMBH is modeled using the \citet{chandr43} formula.  Assuming a Maxwellian velocity distribution, the drag force on a BH moving with velocity $v_{\rm BH}$ through a background density $\rho_*$ is:
\begin{equation}
{\bf f_{\rm df}} = - I(\mathcal{M})  \times  { 4 \pi \rho_* (G M_{\rm BH})^2 \over \sigma_*^2 }   { {\bf v_{\rm BH}} \over v_{\rm BH} },\label{eq:dfs}
\end{equation}
where 
\begin{equation}
I(\mathcal{M}) = {{\rm ln}(\Lambda) \over \mathcal{M}^2} \left ( {\rm erf}\left ({\mathcal{M} \over \sqrt{2}} \right ) -  \sqrt{2 \over \pi} \mathcal{M} e^{-\mathcal{M}^2/2} \right ), 
\end{equation}
and the Mach number $\mathcal{M}\equiv v_{\rm BH} / \sigma_*$.  We use 3.1 for the Coulomb logarithm ln($\Lambda$) \citep{escala04, guamer08}.  

\subsection{Gaseous disk component}
\label{ssec:gasdisk}

In addition to a stellar bulge, most galaxies will contain a significant gaseous component.  This is especially true of merger remnants, as mergers are more frequent between galaxies at high redshift, where gas fractions are generally higher.  We assume that by the time a SMBH binary has merged in the galaxy center, the gas has cooled into a disk.  For our fiducial model, we assume the disk mass is determined by the gas fraction $f_{\rm gas} \equiv M_{\rm disk}/M_{*} = 0.5$, {\rev such that the galaxy is relatively gas-rich}.   

\subsubsection{Disk structure}
\label{sssec:structure}

\begin{figure}
\resizebox{\hsize}{!}{\includegraphics{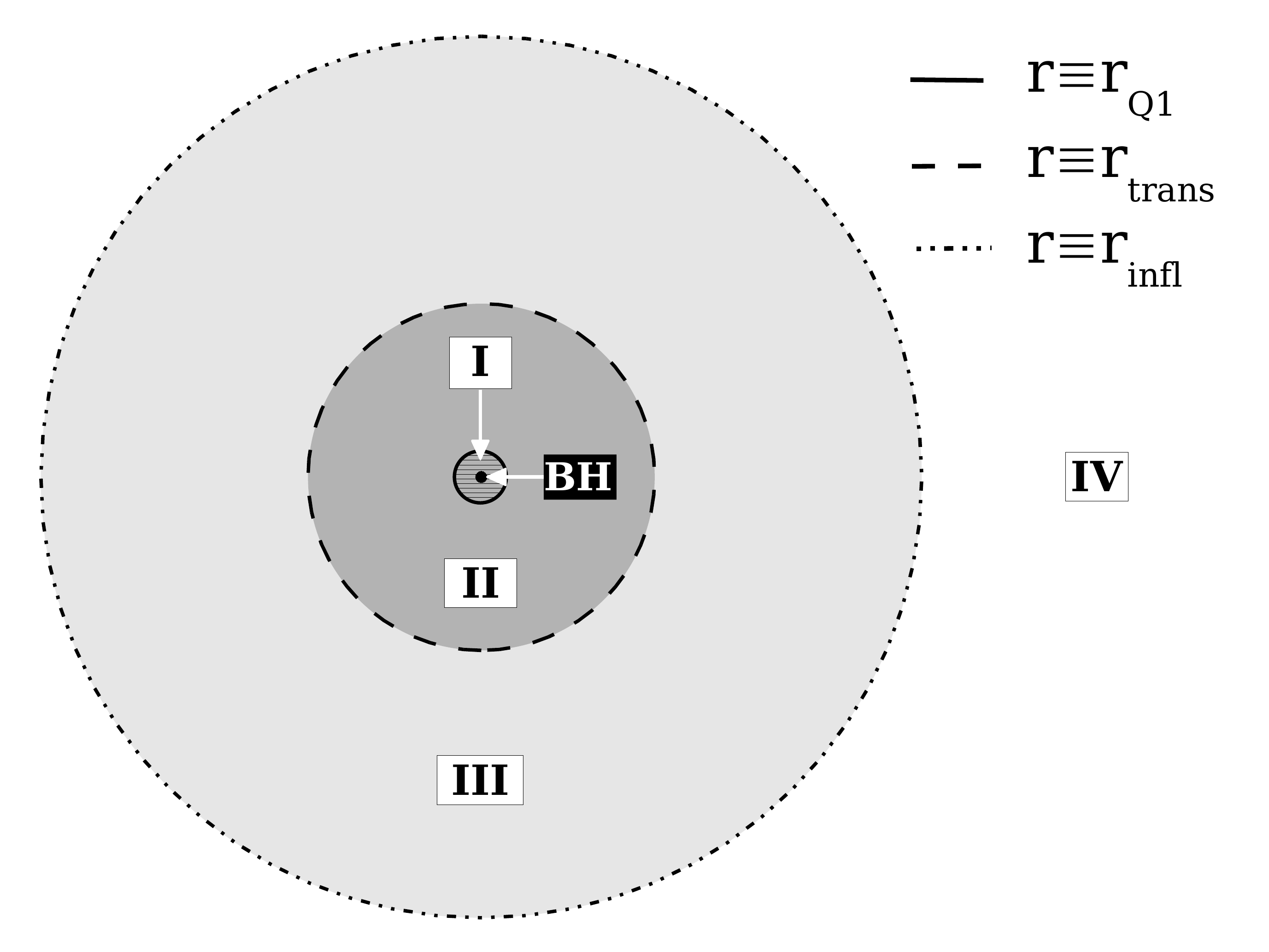}}
\caption[Disk model diagram.]{Schematic face-on diagram of our model for the gaseous galactic disk (not to scale).  Zone I is the standard \citet{shasun73} $\alpha$-disk, marked with horizontal hash-marks.  The solid circle denotes the radius $r_{\rm Q1}$.  Zone II (shown in dark gray) is the inner (BH-dominated) portion of the self-gravitating accretion disk, while Zone III (shown in light gray) is the outer (disk-dominated) portion of the accretion disk.  The transition between the two, $r_{\rm trans}$, is marked with the long-dashed circle.  The BH radius of influence, $r_{\rm infl}$, is marked with the dot-dashed circle.  Beyond this radius is the exponential disk (Zone IV), which has no defined outer boundary. \label{fig:disk}}
\end{figure}

Analytical models of disk structure generally fall into two distinct categories: SMBH-dominated, viscosity-driven accretion disks (i.e., $\alpha$-disks) with $M_{\rm disk} \ll$ $M_{\rm BH}$, and self-gravitating, star-forming galactic disks with $M_{\rm disk} \gg M_{\rm BH}$.  The intermediate regime, $M_{\rm disk} \sim M_{\rm BH}$, is less straightforward to model analytically and is seldom resolved by hydrodynamical simulations.  This regime also defines the SMBH radius of influence, $r_{\rm infl}$, through the relation $M_{\rm disk}(<r_{\rm infl}) = 2M_{\rm BH}$.  \citet{goodma03} \& \citet{gootan04} discuss the issue of self-gravitating accretion disks, though they focus on scales of $\la$ pc where the disk is still nearly Keplerian (as in our Zone II, defined below).  \citet{thomps05} conduct a more detailed analysis of this problem all the way to the BH radius of influence using a semi-analytical model of the disk structure that includes dust; this is a more involved approach than is necessary for the scope of our paper.  

In addition to seeking a simpler disk model, we address the more general question of whether a purely analytic disk solution can be constructed for the entire region between the self-gravity radius and the BH radius of influence.  Solutions have been derived by other authors for various limits, but to our knowledge this is the first attempt to extrapolate these limits to model the entire disk.  Our approach is to construct an analytic model of the disk structure at all radii, from the innermost stable circular orbit around the BH to the outer reaches of the galaxy.  Because this range incorporates vastly different physical environments that cannot be modeled by a single set of simple equations, we define several regimes separated by critical radii.  Fig.~\ref{fig:disk} shows the four concentric ``zones" we have defined in this ``onion-skin" model.  {\rev Extending the disk continuously from galactic scales down to the innermost radii is important for calculating the accretion rate and ejected mass for a wide range of recoil velocities.  However, we note that BHs kicked out of the disk plane spend a small fraction of their wandering time in the inner disk regions, such that the {\it dynamics} of these BHs will generally be dominated by the outermost disk region (Zone IV) and the stellar distribution.  Thus, our dynamical results are largely independent of the details of the inner disk model described here.}

The innermost zone, Zone I, is a standard $\alpha$-disk \citep{shasun73}.  The outer edge of this region, $r_{\rm Q1}$, is defined as the radius at which the disk becomes unstable according to the \citet{toomre64} parameter,
\begin{equation}
 \label{eq:toomre}
Q = { c_{\rm s} \kappa_{\Omega} \over \pi G \Sigma},
\end{equation}
where $c_{\rm s}$ is the gas sound speed, $\kappa_{\Omega}^2 = 4\Omega^2 + r (\partial\Omega^2/\partial r)$ is the epicyclic frequency defined in terms of the rotation angular frequency, $\Omega$, and $\Sigma$ is the disk surface density.  $Q < 1$ corresponds an unstable (i.e., self-gravitating) disk.  Following \citet{gootan04}, we assume radiation pressure dominates gas pressure at $r \la r_{\rm Q1}$, such that the sound speed is simply given by $c_{\rm s}^2 = 4 \sigma_B T^4 / 3 c \rho$, where $\sigma_B$ is the Stefan-Boltzmann constant.  {\rev Although radiation pressure dominates in this region, we assume the viscosity scales only with the gas pressure; otherwise the disk may be unstable.}  From {\rev these assumptions} one can derive the radius $r_{\rm Q1}$ \citep{goodma03, gootan04}:
\begin{eqnarray}
r_{\rm Q1} \approx 1.5\times10^{-2}\, {\rm pc} \left ({\alpha/0.3\over\mu}\right )^{1/3} \left({ f_{\rm Edd}\over \epsilon_{\rm rad}/0.1}\right )^{1/6} \left ({ \kappa\over\kappa_{\rm es} }\right )^{1/2} \\ \nonumber
\times \left({ M_{\rm BH}\over10^8 {\rm M}_{\odot}}\right )^{1/2}. 
\end{eqnarray}
Here $\mu$ is the mean molecular weight of the gas, $f_{\rm Edd} = \dot M_{\alpha}/\dot M_{\rm Edd}$ is the ratio of the $\alpha$-disk accretion rate to the Eddington rate, $\epsilon_{\rm rad}$ is the radiative efficiency, and the opacity $\kappa$ is scaled to the electron-scattering value, $\kappa_{\rm es} = 0.4$.  We use $\kappa = \kappa_{\rm es}$ in our calculations.  For our fiducial parameters, the radius at which the radiation and gas pressures are equal is $r_{\rm peq} \approx r_{\rm Q1}$, and the dependence of r$_{\rm Q1}$ on BH mass is weak; thus, it is reasonable to calculate $r_{\rm Q1}$ assuming that Zone I of the disk is radiation-pressure dominated in general (see Table~\ref{table:models}).  {\rev Likewise, for $r > r_{\rm Q1}$ the disk is assumed to be gas-pressure dominated.} 

The self-gravitating region beyond $r_{\rm Q1}$ will cool and undergo star formation, which will heat the disk via feedback processes.  Gas compression and shocks may also form due to gravitational instabilities created when $Q < 1$.  It has frequently been argued that these feedback and cooling processes will balance, leaving the disk in a marginally stable state with $Q \sim 1$ \citep[e.g.,][]{hohl71, paczyn78, bertin97}.  This is supported by results from various simulations \citep[e.g.,][]{gammie01, lodric04}.  Accordingly, we constrain the disk beyond $r_{\rm Q1}$ to have $Q=1$, which is effectively an equation balancing heating and cooling processes in the gas.  We derive our set of equations using this constraint in place of the equation for balance between radiation and viscous dissipation of orbital energy \citep[see, e.g.,][]{bertin97, berlod99, goodma03}.  We retain the assumption of a steady accretion rate via viscous dissipation of angular momentum,
\begin{equation}
\label{eq:mdotkep}
{\dot M}_{\alpha} = 3 \pi \nu \Sigma = 3 \pi \alpha c_{\rm s} h \Sigma = {\rm const},
\end{equation}
where the viscosity $\nu = \alpha c_{\rm s} h$ is defined according to the $\alpha$-disk prescription, $h$ is the disk scale height, and the dimensionless constant $\alpha$ is a free parameter, usually assumed to be in the range $\sim 0.01 - 0.1$.  \citet{goodma03} suggests that for self-gravitating, $Q = 1$ disks, this value may be as high as 0.3.  This is the fiducial value we use in our disk calculations.   Consumption of gas via star formation in the disk is neglected, such that ${\dot M}$ remains constant, although the energy input from star formation is included indirectly by setting $Q = 1$.  We also use the thin-disk approximation, as well the relation $\rho_{\rm g} \equiv \Sigma / 2 h$ for the gas density. 

We construct the intermediate disk regime as follows.  Just beyond $r_{\rm Q1}$, the BH still overwhelmingly dominates the gravitational potential.  Likewise, at the BH radius of influence, $r_{\rm infl}$, the disk becomes dominated by its own gravitational potential.  We use these two limits to define two zones (Zones II \& III) in the region $r_{\rm Q1} < r < r_{\rm infl}$: in Zone II the disk contribution to the potential is neglected, and in Zone III the BH term is neglected.  The transition radius, $r_{\rm trans}$, is defined where the two density profiles intersect ($\Sigma_{\rm II} = \Sigma_{\rm III}$), thus creating a smooth transition between the two regimes.  In the thin-disk, BH-dominated regime (Zone II), $c_{\rm s} = h \Omega_{\rm BH}$, where $\Omega_{\rm BH}^2 = G M_{\rm BH} / r^3$.  The relation for $c_{\rm s}$, along with Eqs.~(\ref{eq:toomre}) \& (\ref{eq:mdotkep}), can be solved to give expressions for $\Sigma, c_{\rm s}$ and $h$:
\begin{equation}
\Sigma_{\rm II} = {\Omega_{\rm BH} \over \pi} \left ({ {\dot M}_{\alpha} \over 3 \alpha G^2 Q^2} \right )^{1/3} 
\end{equation}
\begin{equation}
c_{\rm s,II} = \left ( { G Q {\dot M}_{\alpha} \over 3 \alpha } \right )^{1/3}
\end{equation}
\begin{equation}
h_{\rm II} = \Omega_{\rm BH}^{-1}  \left ( { G Q {\dot M}_{\alpha} \over 3 \alpha } \right )^{1/3}.
\end{equation}

For Zone III, $c_{\rm s}$ is approximated by using only the disk potential term to solve for vertical hydrostatic equilibrium and the radial Poisson equation, giving $c_{\rm s}^2 = 2 \pi G \Sigma h$.  $\Omega(r)$ similarly is taken to depend only on the disk mass, $M_{\rm disk}$: $\Omega(r)^2 = G M_{\rm disk}(<r) / r^3$.  {\rev This is the form of $\Omega(r)^2$ for a spherical distribution; using this expression for a disk is initially a simplifying assumption.  However, we find a self-consistent solution for a disk with a $1/r$ surface density profile -- i.e., a Mestel disk \citep{mestel63}.  In this special case, the disk's rotation curve exactly equals that of an equivalent spherical distribution, so $\Omega(r)$ is not an approximation.}  The general equation for the viscous mass accretion rate in a non-Keplerian disk is:
\begin{equation}
\label{eq:mdotnonkep}
{\dot M}_{\alpha} = 2 \pi \nu \Sigma \; \vline { {\rm d\, ln} \Omega \over {\rm d\, ln} r} \vline = {\rm const}.
\end{equation}  
Eqs.~(\ref{eq:toomre}) \& (\ref{eq:mdotnonkep}), along with the sound speed equation, yield the disk quantities for Zone III:
\begin{equation}
\Sigma_{\rm III} =   \left ( { 4 \over \pi Q^2 } \right ) \left ( { {\dot M}_{\alpha}^2 \over \alpha^2 G } \right )^{1/3} r^{-1}
\end{equation}
\begin{equation}
c_{\rm s,III} = \left ( { G \dot M_{\alpha} \over \alpha } \right )^{1/3}
\end {equation}
\begin{equation}
h_{\rm III} = { Q^2 \over 8 } r.
\end{equation}

This results in an angular velocity $\Omega(r)^2 = (c_{\rm s,III}/h_{\rm III})^2\, (h_{\rm III}/r)$.  In a recent review, \citet{lodato08} also derives the set of equations for what we call Zone III and defines the radius of transition to a Keplerian disk.  Note that his numerical factors differ slightly from ours due to a different definition of the disk scale height.

We model the region beyond $r_{\rm infl}$ (Zone IV) with an exponential profile, $\Sigma_{\rm IV} = \Sigma_0{\rm e}^{-r/r_{\rm disk}}$, where the scale radius $r_{\rm disk}$ is set by matching $\Sigma_{\rm III}(r_{\rm infl}) = \Sigma_{\rm IV}(r_{\rm infl})$.  $\Sigma_0$ is obtained by integrating $\Sigma$(r) from $r_{\rm infl}$ and equating this to the gas mass beyond $r_{\rm infl}$:
\begin{equation}
\Sigma_0 = { M_{\rm disk}(>r_{\rm infl}) \over 2 \pi r_{\rm disk}^2 \left ( { r_{\rm infl} \over r_{\rm disk} } + 1 \right ) e^{-r_{\rm infl}/r_{\rm disk}}}. 
\end{equation} 

To calculate the disk gravitational potential at points outside the disk plane, we again use the approximation of spherical symmetry for the gravitational potential, namely, $\partial \Phi/\partial R = G M(<R)/R^2$.  However, a perfectly spherically symmetric potential is clearly unphysical given the asphericity of the gas disk.  To break this symmetry, we introduce a small elliptical perturbation to the potential by writing the radial coordinate as $\tilde R = \sqrt{ r^2 + (z/q)^2 }$, where $q<1$ is the axial ratio of an equipotential ellipsoid.  This parameter is not intended to accurately reproduce the ellipticity profile of the galactic potential at all radii -- in fact, in this approximation the ellipticity is constant at all radii.  Instead, it merely allows some precession of the orbits in the $r-z$ plane, as one expects due to torques introduced by the aspherical mass distribution.  As a fiducial value in our runs, we choose $q = 0.99$ (i.e, a 1\% distortion of the axial ratio).  We find that the largest effects due to this perturbation occur for near-escape orbits with $v_{\rm k} \approx v_{\rm esc}$, where the BH reaches the largest radii.  This further validates the choice of small $q$, since in reality the galactic potential will tend toward sphericity at large radii.  See \S~\ref{ssec:q_sensitivity} for further discussion on the sensitivity of our results to this parameter.

When the SMBH is ejected from the center of the galaxy, it carries with it a portion of the gaseous disk out to a radius $r_{\rm ej}$.  This radius can be estimated simply as $r_{\rm ej} = G M_{\rm BH} / (v_{\rm k}^2 + c_{\rm s}^2)$; namely, the gas orbiting the BH with $v_{\rm orb} > v_{\rm k}$ remains bound to the BH as it leaves the galaxy center \citep{loeb07}.  We thus assume the total mass ejected is $M_{\rm ej} = M_{\rm BH} + M_{\rm disk,ej}$.  Another consideration is that numerous studies have found that in hydrodynamical simulations a binary BH with moderate mass ratio (such as would produce large recoils) creates torques on the surrounding gas disk that carve out a circumbinary gap about twice the size of the semimajor axis {\rev \citep{liu03, escala05, milphi05, macmil06, hayasa07}}.  As the binary approaches merger, the gravitational-wave merger timescale, $t_{\rm gw}$, decreases rapidly, and the BH orbit decays faster than the hole's viscous timescale to refill, $t_{\rm visc}$.  The final gap size at merger is determined by the point at which $t_{\rm gw} << t_{\rm visc}$, which for our models is $\sim 100\, r_{\rm S}$ \citep[][$r_{\rm S}$ is the Schwarzschild radius]{milphi05}.  The disk mass within this radius is negligible, so we can ignore the circumbinary gap in our simulations.

We also assume that the ejected disk leaves a hole in the galactic gas disk of radius $r_{\rm ej}$ in our model.  The hole will {\rev eventually refill; we assume refilling will take place on} the viscous timescale $t_{\rm visc} \sim r^2/\nu = r^2/\alpha c_{\rm s} h$, {\rev though we note that \citet{milphi05} have suggested that the hole should refill faster than the viscous timescale.}  We calculate this timescale in each of our runs, based on the environment at the hole radius $r_{\rm ej}$.  For BHs kicked into the disk plane with low velocity, this timescale will be $\ga 10^8$ yr, which in this case is generally longer than $t_{\rm fin}$, the time for the BH to settle back to the center.  For out-of-plane kicks, $t_{\rm visc} > t_{\rm fin}$ for low $v_{\rm k}$, when $r_{\rm ej}$ is large.  The velocity $v_{\rm k}$ where $t_{\rm visc} = t_{\rm fin}$ depends on the galaxy model and kick inclination, and is $\sim 300-400$ km s$^{-1}$ for our fiducial model.  We neglect the refilling of the central hole in these cases.  However, for larger kick velocities, the hole is very small ($r_{\rm ej} \propto 1/v^2$) and makes little difference to the overall dynamics and accretion of the moving BH.    

\subsubsection{Gas dynamical friction}
\label{sssec:gasdf}
When the BH $+$ ejected disk are moving through the galactic disk, gas dynamical friction acts on them in addition to stellar dynamical friction.  \citet{ostrik99} derived an analytical formula for a body moving through a gaseous medium that has the same form as Chandrasehkar's formula, but with a different function I($\mathcal{M}$) and with $\mathcal{M}\equiv v_{\rm BH} / c_{\rm s}$.  This formula provides a fairly accurate estimate of gas drag forces and has been used by many other authors \citep[e.g.,][]{naraya00, karsub01, kim05}.  However, \citet{escala04}, as well as \citet{sanbra01}, have found that Ostriker's formula overestimates the dynamical friction somewhat for slightly supersonic velocities, $\mathcal{M} \sim 1$.  To get around this problem, we adopt the approach of \citet{escala04} and choose a simple parametric form for I($\mathcal{M}$) that approximates Ostriker's formula for a given value of ln($\Lambda$) but lowers the drag force in the region $\mathcal{M} \sim 1 - 2$.  Assuming ln$(\Lambda) = 3.1$ as a reasonable nominal value \citep[cf.][]{lintre83, cora97}, we use Chandrasekhar's formula (Eq.~\ref{eq:dfs}) for the gas dynamical friction, but with ln$(\Lambda) = 4.7$ for $\mathcal{M} \geq 0.8$ and ln$(\Lambda) = 1.5$ for $\mathcal{M} < 0.8$.  

We must also consider that the kinetic energy removed from the SMBH by dynamical friction is added to the gas disk; this energy input is not negligible.  Detailed treatment of this process would require a full hydrodynamical approach, but ignoring it completely yields unphysical results.  We therefore modify our prescription for gas dynamical friction to account for the heating of the disk as follows.  We assume the disk is ``puffed up" {\rev to a} modified scale height $h' \sim r$, such that the gas density is reduced by a factor $h/r$ and the sound speed $c_{\rm s}$ is increased by a factor $(h/r)^{-1}$, where $h$ is the initial scale height.  We furthermore assume, due to the large initial energy input as the BH moves through the cold disk, that this heating occurs effectively instantaneously, and we always use the modified quantities in the dynamical friction calculation.

\subsubsection{SMBH mass accretion rate}
\label{sssec:mdot}
In addition to tracking the trajectory of the SMBH, we also make an estimate of the accretion rate at each timestep.  Nominally, the accretion rate is set to $\dot M_{\alpha}$, a free parameter for $r < r_{\rm infl}$ (disk zones I - III).  We define ${\dot M}_{\alpha} = f_{\rm Edd} \times {\dot M}_{\rm Edd}$, where
\begin{equation}
{\dot M}_{\rm Edd} = {4\pi G M_{\rm BH} \over \kappa_{\rm es\,} c\, \epsilon_{\rm rad}},
\end{equation} 
and we adopt a value of $\epsilon_{\rm rad}=0.1$ for the radiative efficiency.  {\rev We set $f_{\rm Edd} = 1$ for the gas-rich ($f_{\rm gas}=0.5$) models, such that the accretion is Eddington-limited.  For consistency we keep the same ratio $f_{\rm Edd} = 2 f_{\rm gas}$ for the $f_{\rm gas}=0.1$ model (such that $f_{\rm Edd} = 0.2$ in this case).}  If the SMBH is out of the plane of the disk, its accretion rate is $\dot M_{\alpha}$ for $0 < t < t_{\rm accr}$, {\rev where $t_{\rm accr}$ is the accretion timescale $M_{\rm disk, ej}/\dot M_{\alpha}$.  After this time,} we assume the ejected disk has been consumed, and the accretion rate is set to zero.  When the SMBH is in the gas disk, either as it leaves the center or on subsequent passages through the disk (``disk crossings''), the accretion rate may change.  Again, at $t < t_{\rm accr}$, we can nominally assume that the BH carries with it a disk of its own that feeds the BH at a rate $\dot M_{\alpha}$.  However, the BH may pass through regions of the galactic gas disk with a much larger scale height than in its own disk, meaning that the BH may accrete mass from all directions, not just its disk plane {\rev -- i.e., mass in the galactic disk may be swept up by the BH}.  The accretion from above and below is more accurately modeled by Bondi-Hoyle accretion \citep{bonhoy44}:
\begin{equation}
\dot M_{\rm B} = { 4 \pi (G M)^2 \rho_{\rm g,\infty}  \over (v_{\rm rel}^2 + c_{\rm s,\infty}^2)^{3/2} } ,
\end{equation}
where ${\bf v_{\rm rel}(r)} = [\, v_{\rm BH,r}(r) {\bf \hat r}, (v_{\rm BH,\phi}(r) - v_{\rm circ})\, {\bf \hat \bphi}, v_{\rm BH,z}(r) {\bf \hat z}\, ]$, $v_{\rm circ}$ = $v_{\rm circ}(r+r_{\rm ej})$ is the circular velocity at the outer edge of the BH disk, and $\rho_{\rm g,\infty}, c_{\rm s,\infty}$ are also evaluated at this radius.  This can be rewritten in terms of the Bondi radius $r_{\rm B} = 2 G M_{\rm ej} / (v_{\rm rel}^2+c_{\rm s}^2)$ as 
\begin{equation}
\dot M_{\rm B} = \pi \rho_{\rm g,\infty} \sqrt{v_{\rm rel}^2 + c_{\rm s,\infty}^2}\, r_{\rm B}^2 f_{\rm geom},
\end{equation}
where we have added the geometric factor $f_{\rm geom} \equiv {\rm min}(1, h/r_{\rm B})$ to account for the reduced cross-sectional area of accretion if the Bondi radius is larger than the physical height of the disk.  Because the BH also carries with it a disk from which it is accreting at a rate ${\dot M}_{\alpha}$, the accretion cannot be modeled by either the Bondi-Hoyle or $\alpha$ prescriptions alone; we therefore introduce a hybrid prescription.  We first assume that as long as $t < t_{\rm accr}$, the BH accretion rate will not fall below $\dot M_{\alpha}$ while passing through the galactic disk, because it accretes at this rate from its bound disk.  Bondi-Hoyle accretion alone decreases with BH velocity as $1/v_{\rm BH}^3$, but it also varies with environment, which depends on radius.  It is unclear {\em a priori} whether the accretion rate will increase or decrease in this case.  We therefore calculate the accretion rate as follows.  $\dot M_{\rm B}$({\bf r},{\bf v}) for a moving object is related to $\dot M_{\rm B}(0)$, the Bondi rate for a stationary, central object by
\begin{eqnarray}
\label{eq:mdotbv}
\dot M_{\rm B}({\rm \bf r}, {\rm \bf v}) = f \times \dot M_{\rm B}(0), \nonumber \\
f = \left ({ c_{\rm s,\infty}(0) \over \sqrt{c_{\rm s,\infty}^2 + v_{\rm rel}^2} }\right )^3 \times  \left ({\rho_{\rm g,\infty} \over \rho_{\rm g,\infty}(0)}\right ) \times \nonumber \\
\left ( { M_{\rm ej} \over M_{\rm BH} } \right )^2 \times \left ({ f_{\rm geom} \over f_{\rm geom}(0) }\right ).
\end{eqnarray}
For a cold disk rather than an idealized infinite, homogeneous background, the low gaseous sound speed in the disk gives a Bondi radius much larger than the BH radius of influence when the BH is stationary, while in reality the two should be about equal.  To avoid calculating unphysical accretion rates much greater than $\dot M_{\rm Edd}$, we define a fiducial sound speed $c_{\rm s, \infty} \equiv (2 G M / r_{\rm infl})^{1/2}$, where $M = M_{\rm BH}$ for $c_{\rm s,\infty}(0)$ and $M = M_{\rm ej}$ for $c_{\rm s,\infty}$.  These are the sound speeds used in the equation above.  Because we know that the BH will nominally accrete at the $\alpha$-model rate instead of the Bondi rate, we adopt the factor $f$ defined in Eq.~(\ref{eq:mdotbv}), which depends on $c_{\rm s,\infty}$, $v_{\rm rel}$,  $\rho_{\rm g,\infty}$, $M_{\rm ej}$, and $f_{\rm geom}$, and normalize it to ${\dot M}_{\alpha}$.  We then have the accretion rate $\dot M = f_{\rm amp} \times \dot M_{\alpha}$, where $f_{\rm amp} \equiv {\rm max}(f, 1)$ prevents $\dot M$ from falling below $\dot M_{\alpha}$.  As mentioned above, this applies only when $t < t_{\rm accr}$; otherwise, the BH is naked when it passes through the galactic disk and its accretion rate is just $\dot M_{\rm B}$({\bf r}, {\bf v}) as given in Eq.~(\ref{eq:mdotbv}).  

From this approach, and also adding the requirement that the accretion rate be Eddington-limited, we can define five distinct accretion regimes for the BH:
\begin{itemize}
\item{ $|z_{\rm BH}| > h, t \leq t_{\rm accr}:  \dot M = \dot M_{\alpha}$ }
\item{ $|z_{\rm BH}| > h, t > t_{\rm accr}:  \dot M = 0$ }
\item{ $|z_{\rm BH}| \leq h, t \leq t_{\rm accr}, f_{\rm amp} = 1:  \dot M = \dot M_{\alpha}$ }
\item{ $|z_{\rm BH}| \leq h, t \leq t_{\rm accr}, f_{\rm amp}>1: \dot M=${\small min}$(f_{\rm amp} \dot M_{\alpha}, \dot M_{\rm Edd})$ }
\item{ $|z_{\rm BH}| \leq h, t > t_{\rm accr}: \dot M=${\small min}$(\dot M_{\rm B}({\bf r}, {\bf v}), \dot M_{\rm Edd})$ }
\end{itemize}

In practice, we find that the amplification factor $f_{\rm amp}$ is almost never greater than unity; slight amplification occurs in our dry-merger model for the (low-velocity) in-plane kicks.  In all other cases the $v^{-3}$ dependence in the Bondi-Hoyle formula dominates the dependence on the changing gaseous environment, such that any contributions to the mass accretion from above and below the BH-disk plane are negligible.  {\rev In other words, a negligible amount of mass is swept up by the BH.}  We can therefore choose $f_{\rm Edd} = 1$ for our fiducial model, such that $\dot M_{\alpha} = \dot M_{\rm Edd}$, and disregard the amplification factor.  When $t > t_{\rm accr}$, however, $\dot M$ is determined solely by the Bondi-Hoyle formula and can have more interesting behavior as seen in Figs.~\ref{fig:fid_i45v440}, \ref{fig:fid_i45v740}, \& \ref{fig:fgas_i45v1100}.

\subsection{Calculation of SMBH trajectories}
\label{ssec:trajectories}
We integrate the trajectories of SMBHs under the influence of gravity and dynamical friction for a range of kick velocities and angles.  The integration is stopped when one of four criteria is met: 
\begin{itemize}
\item{ The BH escapes from the galaxy.  The actual escape velocity is slightly higher than that calculated from the gravitational potential alone, due to the drag from dynamical friction.  We consider the BH ``escaped" when it exceeds the nominal escape velocity, $\sqrt{-2\Phi}$, and is still moving away from the galaxy at a distance of 5 $r_{\rm disk}$. }
\item{ The BH settles back in the center of the galaxy, with velocity $v_{\rm BH} < 0.01 \sigma_*$, where $\sigma_* = \sigma_*(R_{\rm e})$ is the stellar velocity dispersion calculated from the $M_{\rm BH} - \sigma_*$ relation. }
\item{ Sufficient energy is injected into the gas via dynamical friction to effectively unbind the disk.  The binding energy of the gas disk is estimated to be $\sim 0.5\, G\, M_{\rm disk} / (5\, r_{\rm disk})$.  The cumulative energy injected by dynamical friction is the sum of the energy at each timestep d$t$, $\sim M_{\rm ej}\, {\rm f_{df}}\, v_{\rm BH}\, {\rm d}t$.  (This limit is not reached in any of our simulations.) }
\item{ The integration time reaches a Hubble time.  (This limit is not reached in any of our simulations.) }\end{itemize}

\begin{table}
\begin{center}
\begin{tabular}{llcrrrr}
\hline
{\bf \large Model} & & & {\bf \large A} & {\bf \large B} & {\bf \large C} & {\bf \large D} \\ 
\hline
{\bf $M_{\rm BH}$} & {\bf [M$_{\odot}$]} & & {\bf 10$^8$} &{\bf 10$^6$} & {\bf  10$^9$} &{\bf 10$^9$} \\ \\
{\bf $f_{\rm gas}$} & & & {\bf 0.5} & {\bf 0.5} & {\bf 0.5} & {\bf 0.1} \\ 
\hline
$M_*$ & [$10^{10} {\rm M}_{\odot}$] & & $4.1$ & ${\rev 0.034}$ & $45$ & $45$ \\ \\
$r_{\rm peq}$ & [$10^{-2}$ pc] & & $4.0$ & $0.026$ & $50$ & $15$ \\ \\
$r_{\rm Q1}$ & [$10^{-2}$ pc] & & $1.5$ & $0.15$ & $4.7$ & $3.6$ \\ \\
$r_{\rm trans}$ & [pc] & & $13$ & $2.8$ & $28$ & $82$ \\ \\
$r_{\rm infl}$ & [pc] & & $60$ & $6.9$ & $160$ & $250$ \\ \\
$r_{\rm disk}$ & [$10^2$ pc] & & $7.8$ & $1.1$ & $20$ & $18$ \\ \\
$a$ & [$10^2$ pc] & & $7.2$ & ${\rev 0.60}$ &  $25$ &  $25$ \\ \\
$\dot M_{\rm Edd}$ & [M$_{\odot}$ yr$^{-1}$] & & $2.2$ & $0.022$ & $22$ & $22$ \\ \\
$\dot M_{\alpha}$ & [M$_{\odot}$ yr$^{-1}$] & & $2.2$ & $0.022$ & $22$ & $4.4$ \\
\hline
\end{tabular}
\end{center}
\caption{Parameters used in different galaxy models.  The free parameters, $M_{\rm BH}$ and $f_{\rm gas}$, are shown in bold.  The other parameters are derived from these as defined in the text. \label{table:models}}
\end{table}

\subsection{Fiducial model parameters}
\label{ssec:fid_params}
In our fiducial model (Model A), we use a BH mass of $10^8$ M$_{\odot}$ and a gas fraction of $f_{\rm gas} = 0.5$.  Other galactic parameters are scaled to these two input parameters as outlined in \S~\ref{ssec:stellar} \& \ref{ssec:gasdisk}, and Table~\ref{table:models} summarizes these quantities.  We test kick inclinations of $i = 0^{\rm o}$, $45^{\rm o}$, and $90^{\rm o}$ relative to the disk plane\footnote{\rev All kicks velocities are oriented along the positive $x$-axis; thus, inclined kicks are initially in the $x$-$z$ plane and in-plane kicks are initially purely on the $x$-axis.}.  {\rev Two merging BHs that have no spin or aligned spins oriented along the binary orbital rotation axis produce a recoil kick in the binary orbital plane.  This very specific configuration would be important only if the BH spins are preferentially aligned by some mechanism, such as torques from the gas disk \citep{bogdan07}.  In this case we can assume that the orbital plane and resultant recoil kick would both be aligned with the disk plane.  Simulations using full numerical relativity have shown that the maximum recoil kick velocity in this case is $\sim 200$ km s$^{-1}$ \citep{gonzal07a}, so we include only data with $v_{\rm k} \leq 200$ km s$^{-1}$ for in-plane ($i = 0^{\rm o}$) runs.} 

{\rev In the absence of empirical evidence for preferentially aligned spins, we must also consider that the distribution of merging BH spins may be closer to random, producing a random distribution of kick inclinations \citep[cf.][]{schbuo07, campan07a, baker08}.  Thus, we also consider kicks with $i = 45^{\rm o}$ \& $i = 90^{\rm o}$.}  For these out-of-plane kicks, recoil velocities could be {\rev very large \citep{campan07a, gonzal07b} -- up to 4000 km s$^{-1}$ \citep{campan07b}.  However,} trajectories are uninteresting for $v_{\rm k} > v_{\rm esc}$, {\rev so we} calculate trajectories for kick velocities ranging from 100 km s$^{-1}$ up to the escape velocity, in increments of 20 km s$^{-1}$.

\subsection{Additional models}
\label{ssec:other_params}
We also test three models in addition to our fiducial model (see Table~\ref{table:models}).  We calculate the same set of trajectories with BH masses of $10^6$ and $10^9$ M$_{\odot}$ (Models B \& C, respectively), representing the lower and higher ends of the observed SMBH mass function.  10$^6$ M$_{\odot}$ BHs are of particular interest because the mergers of these BHs would be observable with the LISA gravitational wave detector\footnote{Website: http://lisa.nasa.gov/}.  In addition, {\rev because models A, B, \& C are "gas-rich" models,} we {\rev also} test a model with $M_{\rm BH} = 10^9$ M$_{\odot}$ and a gas fraction of $f_{\rm gas} = 0.1$ (Model D), corresponding to a massive, low-$z$ galaxy that has undergone a nearly ``dry" merger.

\section{Results}
\label{sec:results}

\subsection{Fiducial model}
\label{ssec:fidmod}

\begin{figure}
\resizebox{\hsize}{!}{\includegraphics{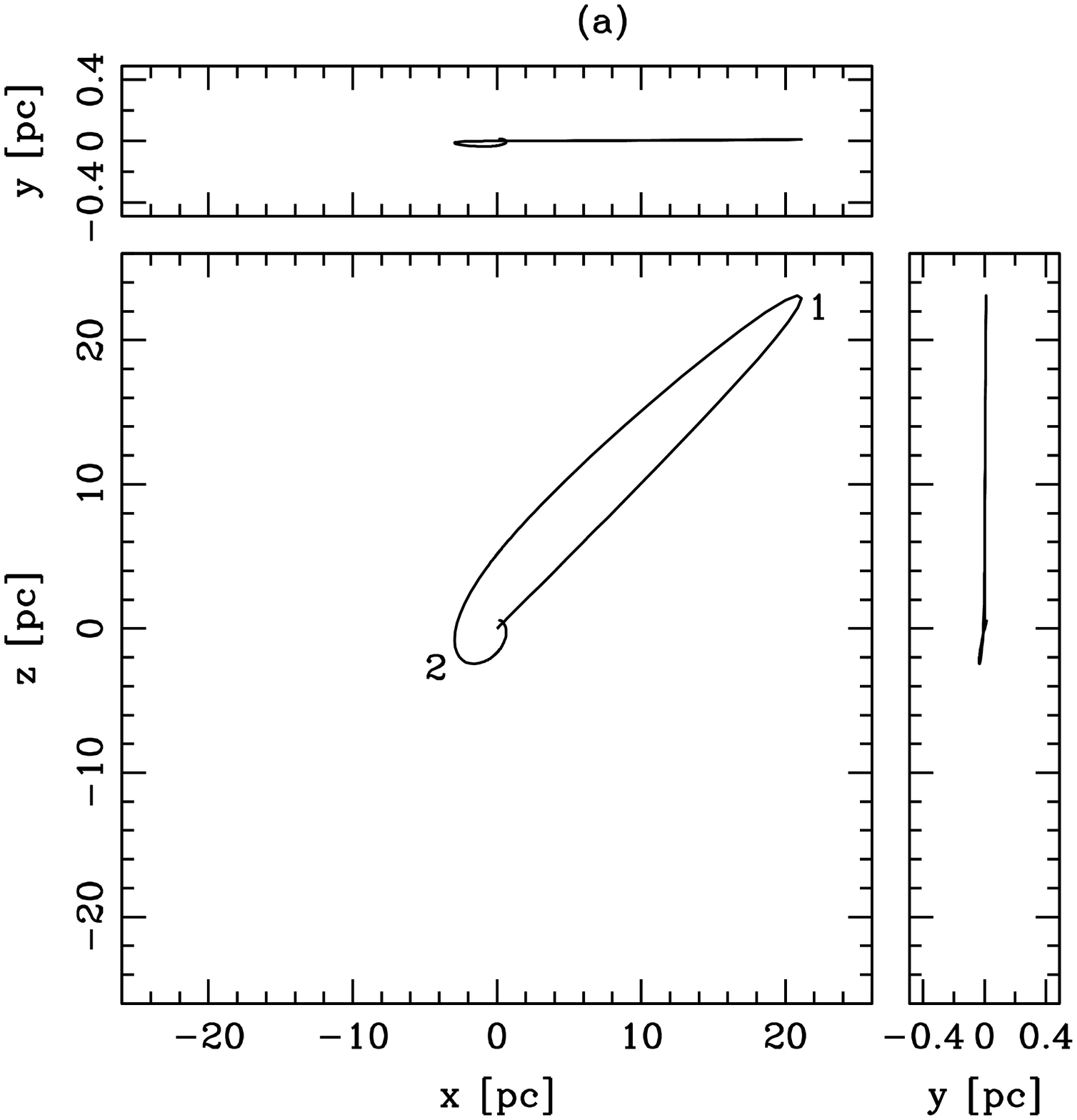}}
\resizebox{\hsize}{!}{\includegraphics{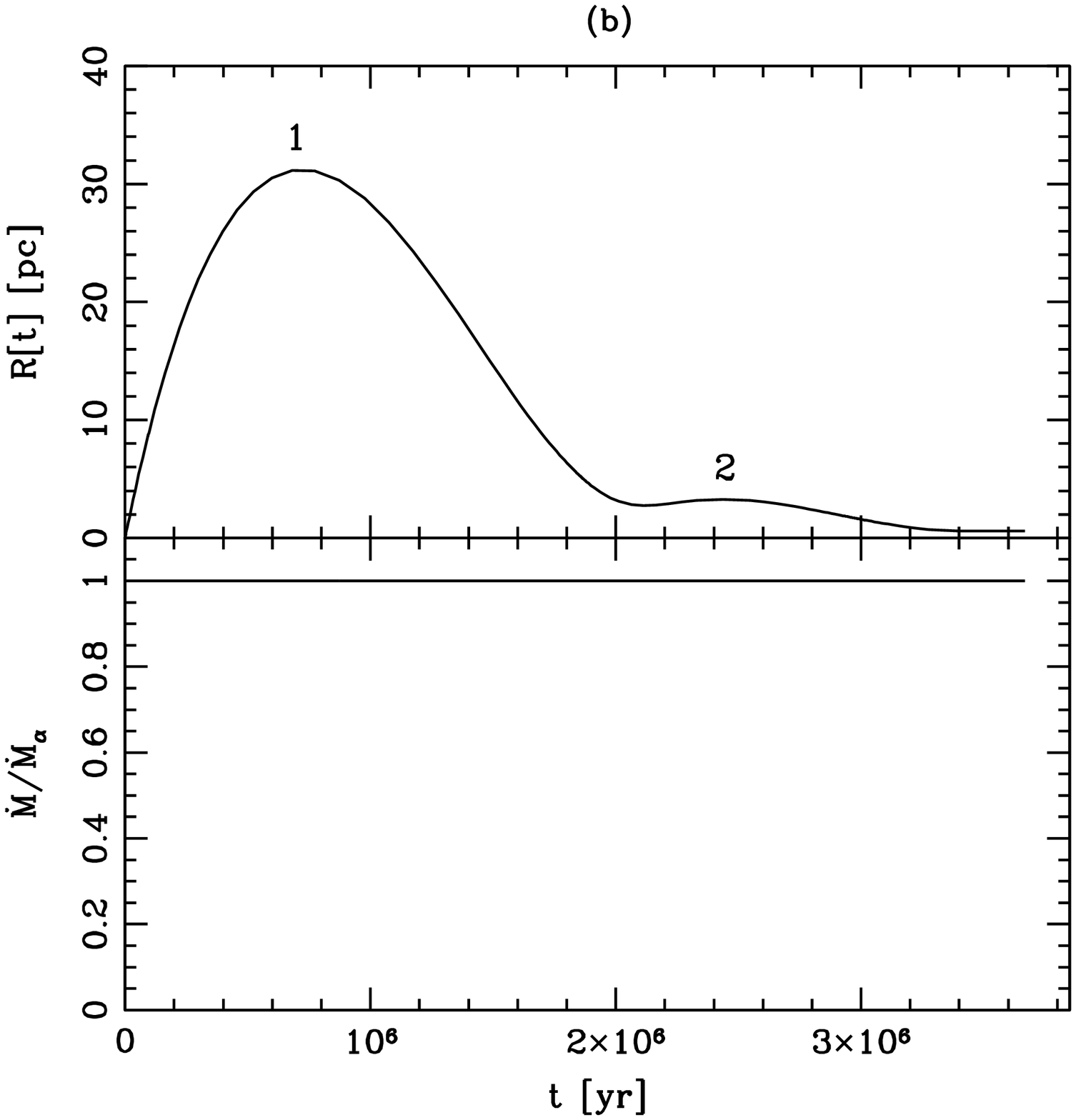}}
\caption[Fiducial model, vk=100, i=45]{\label{fig:fid_i45v100}  {\it (a)} SMBH trajectory for Model A run with $i = 45^{\rm o}, v_{\rm k} = 100$ km s$^{-1}$, seen from three orthogonal orientations.  {\rev Turnaround points are numbered in sequence to show the BH trajectory.}  The $y$-axis (top \& side panels) has been expanded by a factor of 10 to show greater detail.  {\it (b)} {\it top panel:} radial distance of SMBH from center vs. time.  {\rev Maxima are numbered to correspond with turnaround points in {\it (a)}.}  {\it Bottom panel:} Calculated mass accretion rate vs. time, normalized  to the $\alpha$-model accretion rate. Note that in this low-$v_{\rm k}$ example, the accretion rate is constant ($\dot M = \dot M_{\alpha}$) for the entire run.}
\end{figure}

\begin{figure}
\resizebox{\hsize}{!}{\includegraphics{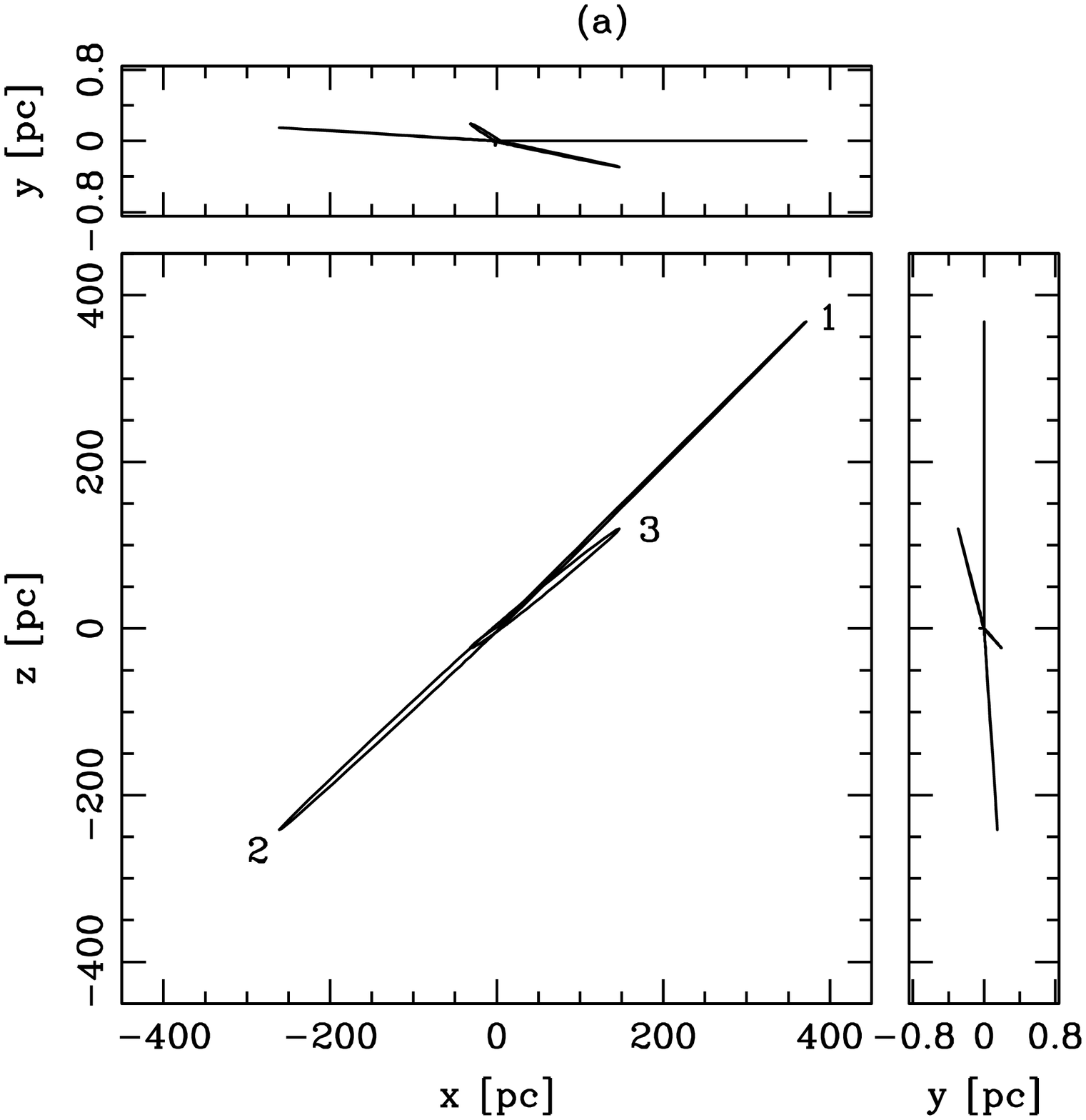}}
\resizebox{\hsize}{!}{\includegraphics{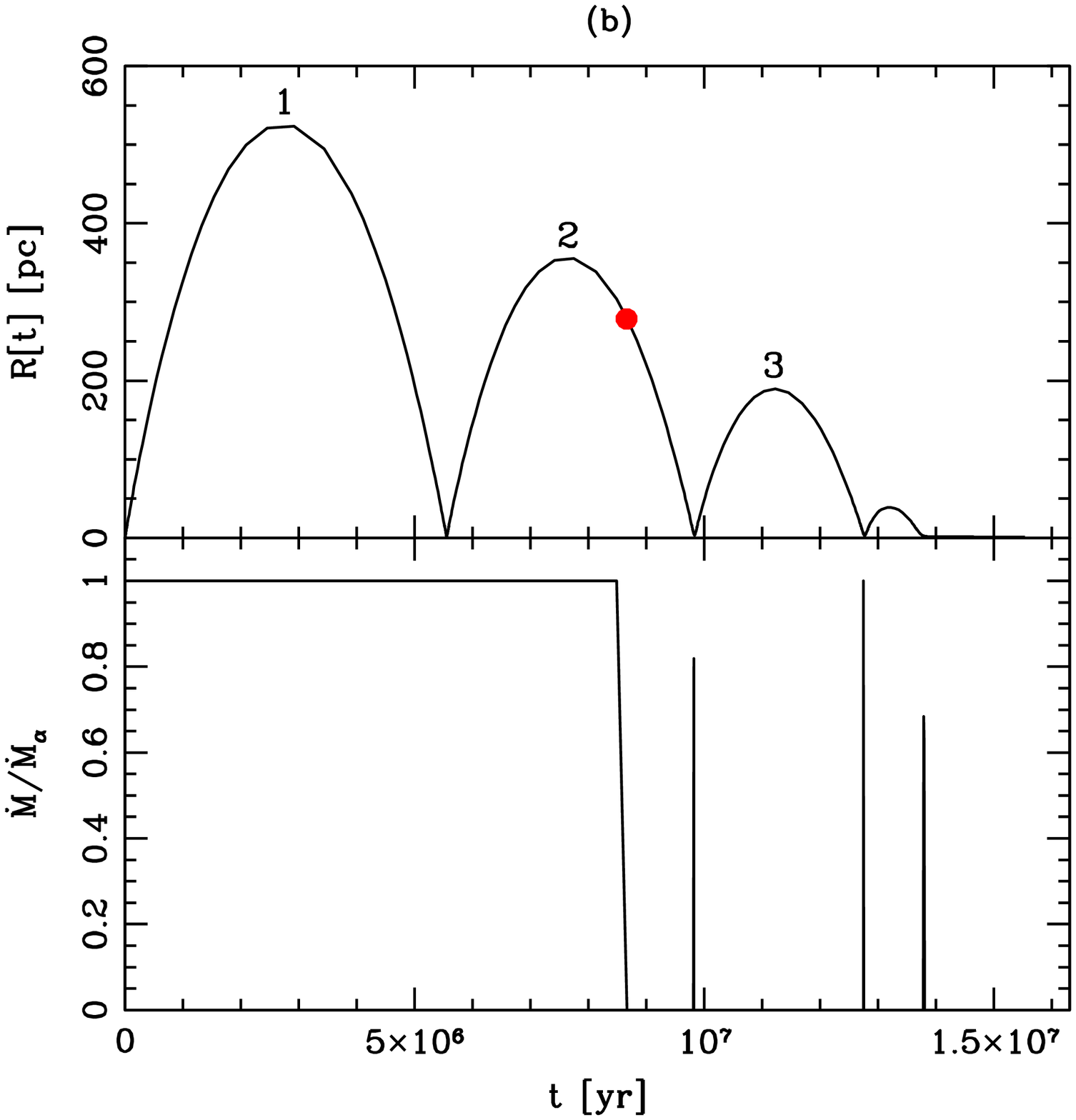}}
\caption[Fiducial model, vk=440, i=45]{ Same data as Fig.~\ref{fig:fid_i45v100}, but with $v_{\rm k} = 440$ km s$^{-1}$.  In Fig.~\ref{fig:fid_i45v440}a, the $y$-axis (top \& side panels) has been expanded by a factor of 100 to show greater detail.  {\rev Turnaround points are numbered in sequence to show the BH trajectory; only the first 3 are numbered.}  The red dot in the top panel of Fig.~\ref{fig:fid_i45v440}b indicates the time $t_{\rm accr} = M_{\rm disk, ej}/\dot M_{\alpha}$ when the ejected disk has been consumed by the SMBH.  Correspondingly, the accretion rate (bottom panel) drops to zero here, and is nonzero again only when the SMBH crosses the galactic disk.\label{fig:fid_i45v440}}
\end{figure}

All the results in this subsection pertain to our fiducial model (Model A).  As explained in \S~\ref{sec:model}, only low velocities ($\la 200$ km s$^{-1}$) are important for in-plane ($i = 0^{\rm o}$) kicks.  Very little happens in these runs; the kick is effectively a small perturbation that is quickly damped out in $\la 10^{6.5}$ yr.  We do not expect these low-velocity, in-plane kicks to be easily observable.

Inclined kicks with low kick velocity are also mostly uninteresting; an example is shown in Fig.~\ref{fig:fid_i45v100}.  Here, the trajectory and mass accretion rate are plotted for the $i = 45^{\rm o}$, $v_{\rm k} = 100$ km s$^{-1}$ run.  The BH (and its accompanying disk) wander about 20 pc away from the galactic center in the $x$ and $z$ directions, settling back to the center in $< 4$ Myr after just one orbital period.  The timescale for the BH to accrete the ejected disk, $t_{\rm accr}$, exceeds the wandering time; hence, the accretion rate $\dot M = \dot M_{\alpha}$ is constant throughout the simulation.  

We have plotted an example of a BH with a moderate kick velocity ($v_{\rm k} = 440 {\rm km\, s}^{-1}$) in Fig.~\ref{fig:fid_i45v440}.  Compared to the low-velocity example in Fig.~\ref{fig:fid_i45v100}, the wandering time of the BH is much longer, $\sim 15$ Myr, and and the BH travels much further from the galactic center.  Note also that the size of the ejected disk, and hence $t_{\rm accr}$, are smaller for higher kick velocities.  In this case, $t_{\rm accr} \approx 9$ Myr is marked with a red dot in Fig.~\ref{fig:fid_i45v440}b, and corresponds to the point at which $\dot M$ first drops to zero.  The subsequent spikes correspond to passages through the disk, or disk crossings, where the BH is again fed by the gas via Bondi-Hoyle accretion.

\begin{figure}
\resizebox{\hsize}{!}{\includegraphics{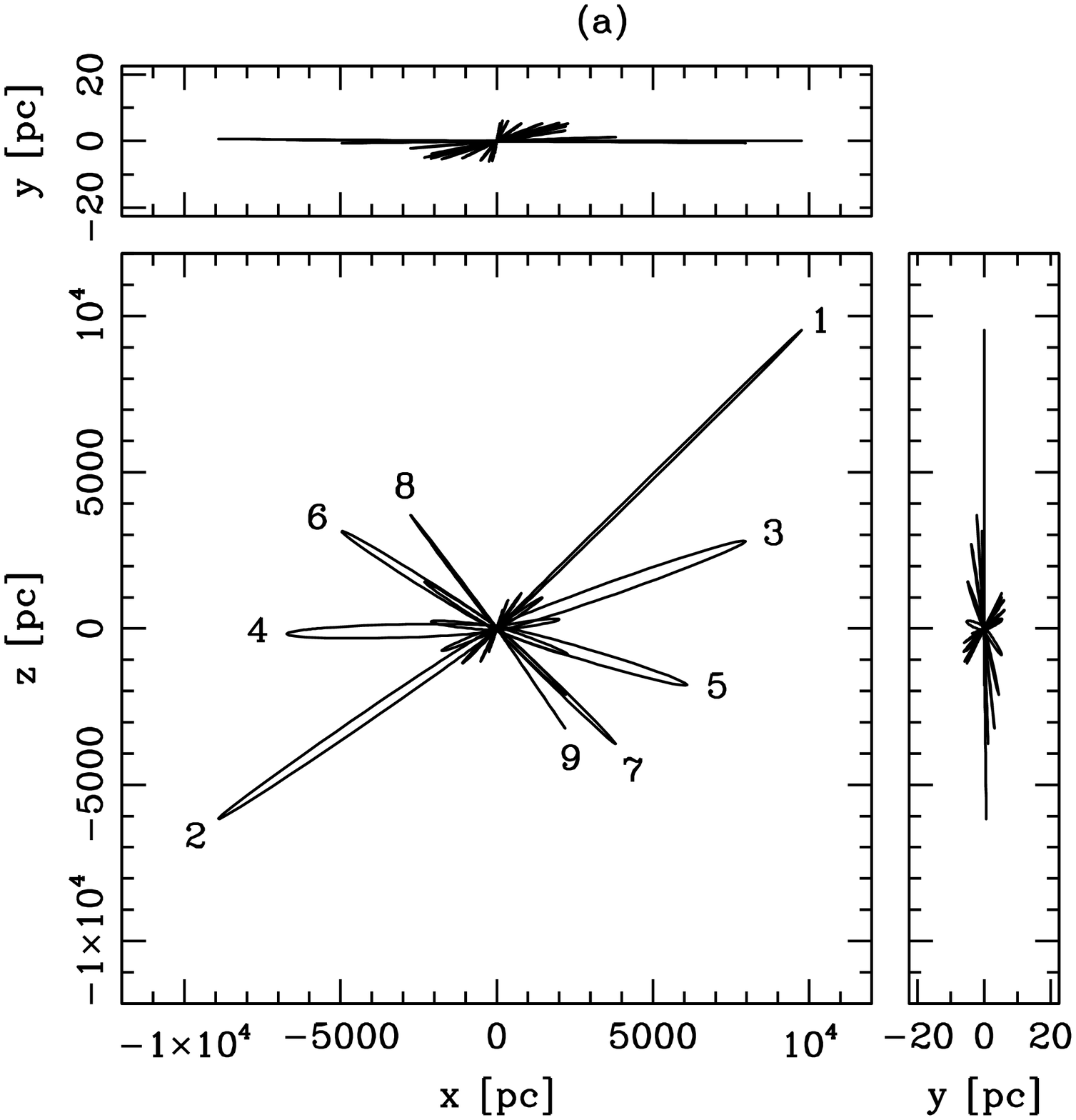}}
\resizebox{\hsize}{!}{\includegraphics{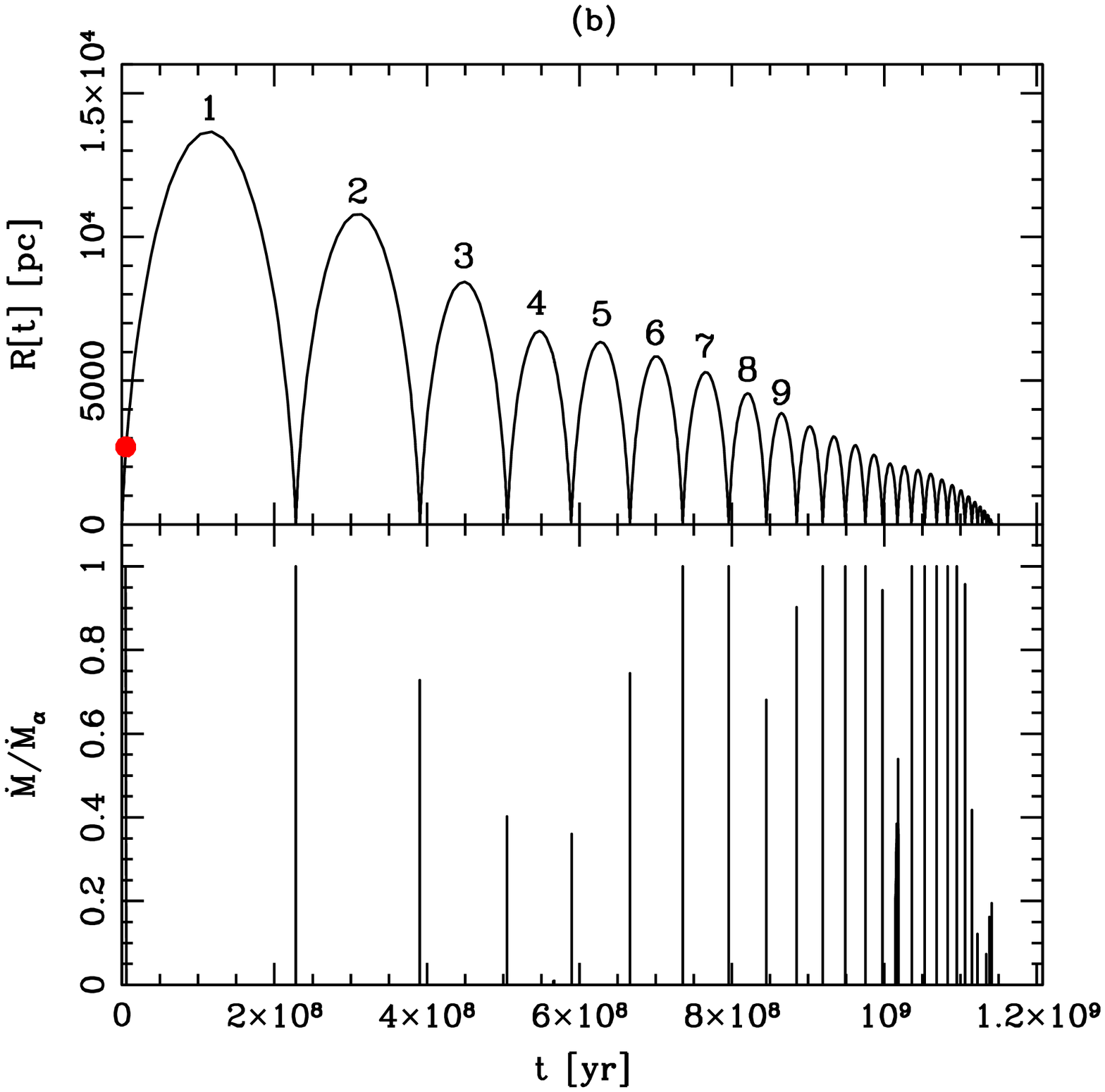}}
\caption[Fiducial model, vk=740, i=45]{Same data as Fig.~\ref{fig:fid_i45v100}, but with $v_{\rm k} = 740$ km s$^{-1}$ (i.e., just below the escape speed for this inclination from this galaxy model).  In Fig.~\ref{fig:fid_i45v740}a, the $y$-axis (top \& side panels) has been expanded by a factor of 100 to show greater detail.  {\rev Turnaround points are numbered in sequence to show the BH trajectory; only the first 9 are numbered.}  The red dot indicates the timescale $t_{\rm accr}$, as described in Fig.~\ref{fig:fid_i45v440}.\label{fig:fid_i45v740}}
\end{figure}

The effective escape speed for $i = 45^{\rm o}$ kicks is between 740-760 km s$^{-1}$.  At $v_{\rm k} = 740$ km s$^{-1}$ (Fig.~\ref{fig:fid_i45v740}), the behavior is a qualitatively similar but greatly amplified version of the moderate-velocity example in Fig.~\ref{fig:fid_i45v440}.  The orbital precession induced by the ellipticity of the galactic potential is quite apparent in this case.  The BH consumes its small ejected disk almost immediately and reaches a maximum distance of $\sim 14$ kpc from the center.  During the $\sim 1.2$ Gyr it takes to settle back to the center, it crosses the disk 28 times, producing multiple short bursts of accretion (Fig.~\ref{fig:fid_i45v740}b).  This is an idealized scenario, however, because on scales $\ga$ kpc, the DM halo potential will start to dominate that of the baryonic galaxy component.  Because the DM distribution is likely to be significantly triaxial on these scales, the BH may {\rev be scattered onto an orbit that takes} significantly longer to return to the center of the galaxy, or it may wander indefinitely in the halo.  \citep[See][for a more detailed analysis of this regime.]{vicari07}  This means that the number of disk crossings and total mass accretion are upper limits, and the BH wandering time is a lower limit.  The amount of accretion will in reality be reduced even further; 1.2 Gyr is well in excess of typical starburst timescales \citep[$\sim 100$ Myr, cf. e.g.,][]{spring05, marcil06}, after which less gas will be left to accrete.  {\rev Additionally, this wandering time is roughly comparable to typical timescales between major galaxy mergers, so this may be another complicating factor.  We note, however, that the upper limit of $\sim 1$ Gyr for the wandering timescale in massive galaxies is a similar result to that of \citet{guamer08}, despite the substantially different physics and parameters that were included in our respective models.}

When the BH is kicked directly out of the plane ($i = 90^{\rm o}$), the effective escape velocity is slightly lower than in the $i = 45^{\rm o}$ case ($\sim$ 660 km s$^{-1}$ for Model A).  The main difference in this case, however, is that very little accretion occurs after $t_{\rm accr}$.  This is because the BH precesses very little about the galaxy center when it is ejected directly along the vertical axis; all of its disk crossings occur very close to the center where the disk height is small, so the disk crossing time is typically too short for any measurable accretion to occur.

\begin{figure}
\resizebox{\hsize}{!}{\includegraphics{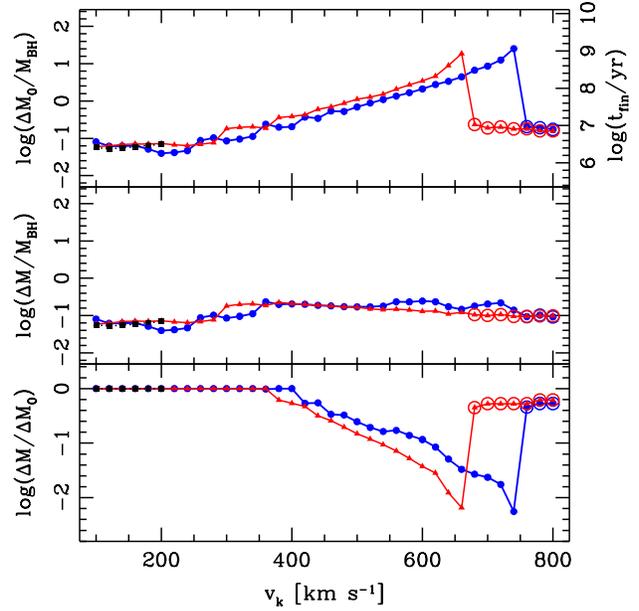}}
\caption[Fiducial model, SMBH growth]{SMBH mass increment for Model A, as a function of kick velocity.  In all panels, black filled squares with dotted line denote $i = 0^{\rm o}$ kicks ({\rev low v$_{\rm k}$ only}), blue filled circles with thick line denote $i = 45^{\rm o}$ kicks, and red filled triangles with thin line denote kicks with $i = 90^{\rm o}$.  Open circles around a point indicate that the BH escaped from the galaxy's baryonic component.  {\it Top panel:} The estimated BH growth, normalized to the initial BH mass, if the BH were to remain in the galactic center for the simulation time ($\Delta M_0 = \dot M_{\alpha} \times t_{\rm fin}$).  Because $\Delta M_0$ scales linearly with $t_{\rm fin}$, the latter is shown on the right vertical axis.  {\it Middle panel:} The estimated BH growth, normalized to the inital BH mass, calculated based on the moving BH's trajectory.  {\it Bottom panel:} The ratio of these two quantities. \label{fig:fid_dM}}
\end{figure}

To understand how GW recoil may affect the mass accretion history of SMBHs, we calculate the total mass accreted by the SMBH during each simulation ($\Delta M$).  We are also interested in how this compares with the mass that would have been accreted by a stationary BH of the same mass over the same time ($\Delta M_0 \equiv t_{\rm fin}/\dot M_{\alpha}$).  Fig.~\ref{fig:fid_dM} shows the SMBH growth as a function of kick velocity.  $\Delta M_0$, and correspondingly $t_{\rm fin}$, increase monotonically with kick velocity until $v_{\rm esc}$ is reached.  In the middle panel where $\Delta M$ is shown, one can see the remarkable result that {\it the mass accreted by wandering SMBHs is fairly constant at $\sim 10\%$ for all kick velocities $\la v_{\rm esc}$}.  This appears to be a coincidental balance between the competing factors of the wandering time, which increases with $v_{\rm k}$, and the time to accrete the ejected disk, which decreases with higher $v_{\rm k}$.  \S~\ref{ssec:obssig} gives further details about this result.

\subsection{SMBH mass models}
\label{ssec:massmod}

\begin{figure}
\resizebox{\hsize}{!}{\includegraphics{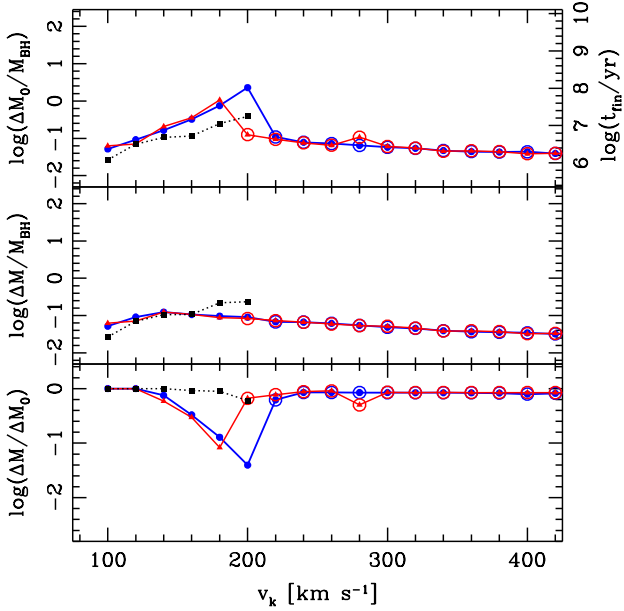}}
\caption[$M_{\rm BH} = 10^6$ M$_{\odot}$ model, SMBH growth]{SMBH growth for Model B as a function of kick velocity.  Notation is the same as in Fig.~\ref{fig:fid_dM}. \label{fig:M6_dM}}
\end{figure}

In Model B with a $10^6$ M$_{\odot}$ BH, kick velocities as low as $\sim 200$ km s$^{-1}$ result in escape of the BH from the galaxy.  Since the magnitude of GW recoil depends on the mass ratio of the progenitor BHs but not on their absolute masses, the probability of large recoil kicks is the same for smaller SMBHs as it is for the most massive ones.  Note that $\Delta M/M_{\rm BH} \sim 10\%$ for $v \la v_{\rm esc}$ in this model as well.  For $v > v_{\rm esc}$, however, one can see that $\Delta M/M_{\rm BH}$ gradually declines, because $t_{\rm accr}$ is also decreasing (see Fig.~\ref{fig:onlytaccr}).

\begin{figure}
\resizebox{\hsize}{!}{\includegraphics{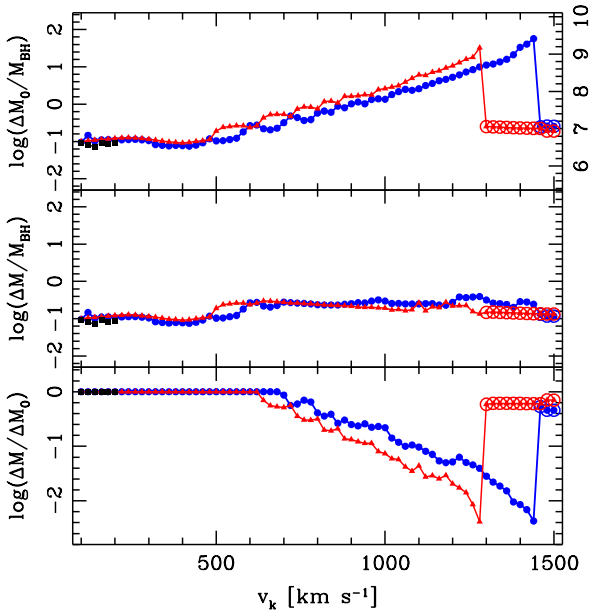}}
\caption[$M_{\rm BH} = 10^9$ M$_{\odot}$ model, SMBH growth]{SMBH growth for Model C as a function of kick velocity.  Notation is the same as in Fig.~\ref{fig:fid_dM}. \label{fig:M9_dM}}
\end{figure}

Fig.~\ref{fig:M9_dM} shows that the growth for a 10$^9$ M$_{\odot}$ BH is qualitatively similar to that of a 10$^8$ M$_{\odot}$ BH, but here the trends are even more apparent.  AGN feedback or gas depletion via star formation should reduce the accretion onto stationary BHs over the long timescales shown in the top panel; the projected growth factor of almost 100 is clearly unphysical.  However, the middle panel demonstrates that recoil kicks are effective at suppressing the BH growth; the growth factor is again remarkably constant at $\sim {\rm few} \times10\%$ regardless of kick velocity.  

\begin{figure}
\resizebox{\hsize}{!}{\includegraphics{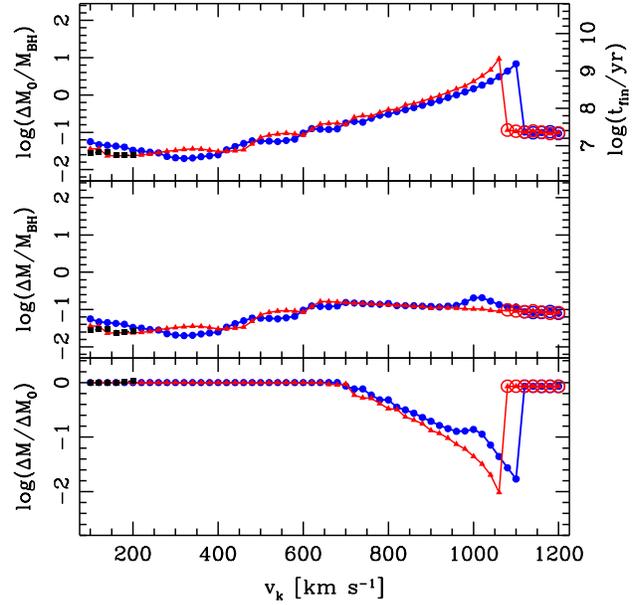}}
\caption[$f_{\rm gas} = 0.1$ model, SMBH growth]{SMBH growth for Model D as a function of kick velocity.  Notation is the same as in Fig.~\ref{fig:fid_dM}. \label{fig:fgas_dM}}
\end{figure}

\subsection{Dry merger model}
\label{ssec:fgasmod}

Model D (a ``dry merger") has the same parameters as the $10^9$ M$_{\odot}$ model above, but with a gas fraction of 0.1.  Fig.~\ref{fig:fgas_dM} shows the BH growth for this model, which is again qualitatively similar to Fig.~\ref{fig:M9_dM}.  As would be expected, the escape velocity is lower for the galaxy containing less gas (note that we have defined $f_{\rm gas}$ such that this model, with lower $f_{\rm gas}$, is less massive overall than the others).  Aside from this, the growth factor for the wandering BH is still $\sim 10\%$, suggesting again that dynamics, not the details of the gas physics, dominate the determination of the final BH mass when the BH is not stationary.  

\begin{figure}
\resizebox{\hsize}{!}{\includegraphics{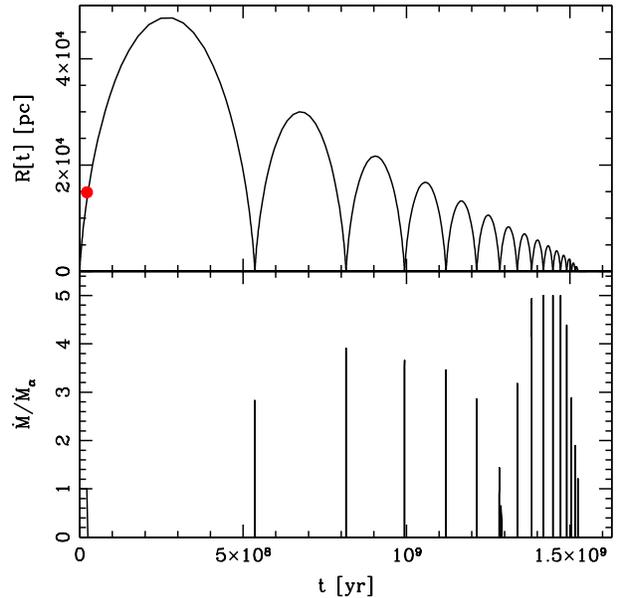}}
\caption[$f_{\rm gas} = 0.1$ model, vk=1100, i=45]{Same notation as Fig.~\ref{fig:fid_i45v100}b, here showing data from Model D for a trajectory with $v_{\rm k} = 1100$ km s$^{-1}$, $i = 45^{\rm o}$.  In this case, Bondi-Hoyle accretion amplifies $\dot M$ above $\dot M_{\alpha} = 0.2 \dot M_{\rm Edd}$ up to the Eddington limit. \label{fig:fgas_i45v1100}}
\end{figure} 
 
Fig.~\ref{fig:fgas_i45v1100} shows an example of the BH accretion in Model D, with $\dot M_{\alpha} < \dot M_{\rm Edd}$.  The kick velocity $v_{\rm k} = 1100$ km s$^{-1}$ is just below the effective escape velocity for this model, so $t_{\rm accr}$ is very short and $\dot M$ quickly drops from $\dot M_{\alpha}$ to zero.  The accretion rate on all subsequent passages through the disk is set by the Bondi rate $\dot M_{\rm B}$, which varies depending on the velocity and position of the disk crossing.  Near the end of the simulation, as the BH slows down, its accretion becomes Eddington-limited.

\subsection{Observational signatures}
\label{ssec:obssig}

Because GW recoil has not been unambiguously observed in any astrophysical system, an important application of this work is to consider the observational signatures that recoiling BHs could produce such that their existence could be directly confirmed.  An observable recoiling BH, i.e. one with significant accretion luminosity, could be distinguished from a stationary central BH via offsets in either physical space or velocity space.  A BH with a resolvable spatial offset from its host galaxy is statistically more likely to have a relatively low velocity, since a large fraction of orbital time is spent at turnaround.  Likewise, the largest velocity offsets will occur either soon after the recoil event or on subsequent passages of the BH through the galactic disk, where the spatial offset from the galactic center is small.  The duty cycle for the former type of event depends on the size of the accretion disk that the recoiling BH carries with it; higher recoil speeds result in smaller ejected disks and shorter duty cycles.  After this disk is exhausted, the most plausible way for the wandering BH to become observable again is by accreting matter as it makes subsequent passages through the galactic disk, {\rev possibly generating a quasi-periodic signature such as knotted or twisted jets}.  Accordingly, we distinguish between two types of ``recoiling quasars:" the former, {\rev primarily} spatially-offset active BHs are identified as ``off-center quasars," and the latter, {\rev primarily} velocity-offset sources are identified as ``disk-crossing quasars."  {\rev Note, however, that there is not a one-to-one correspondence between quasar velocity offsets and disk-crossings, nor between spatial offsets and ``off-center" quasars.  In particular, ejected quasars will often only be visible for a short time after ejection, while they still carry an accretion disk, so that off-center quasars will often have a significant velocity offset as well.  Similarly, many ``disk-crossing" events occur when the BH is settling back to the center of the galaxy and slowing down, so their velocity offsets will be small.}  Note {\rev also} that the term ``quasar" is used loosely here to refer to a source with observable accretion activity; recall that in our runs all BHs except those in Model D radiate at their Eddington luminosities.

\begin{figure}
\resizebox{\hsize}{!}{\includegraphics{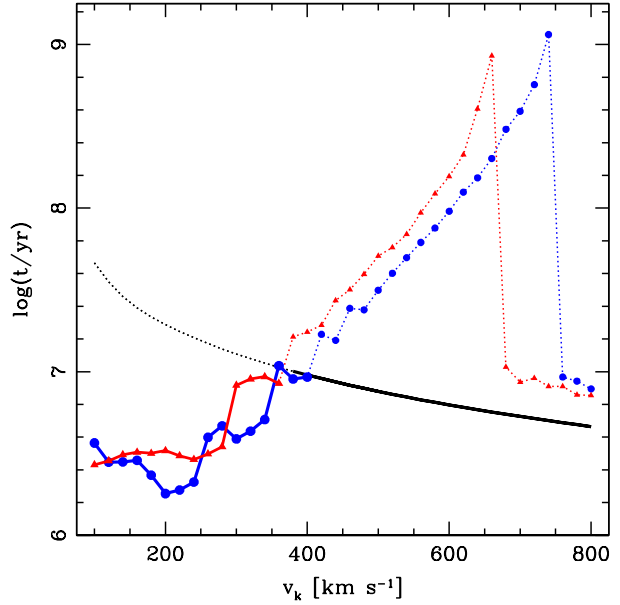}}
\caption[Fiducial model, $t_{\rm accr}$ and $t_{\rm fin}$ versus $v_{\rm k}$.]{Relevant timescales for off-center quasars.  {\it Black line:} lifetime $t_{\rm accr} = M_{\rm disk,ej}/\dot M_{\alpha}$ of the accretion disk ejected with the BH.  {\it Blue line with filled circles:} simulation time $t_{\rm fin}$ for runs with $i = 45^{\rm o}$.  {\it Red line with filled triangles:} $t_{\rm fin}$ for runs with $i = 90^{\rm o}$.  In all cases, the bold portions of the line indicate the minimum of ($t_{\rm accr}$, $t_{\rm fin}$) at each $v_{\rm k}$; remaining portions are shown with a dotted line.\label{fig:fid_taccr}}
\end{figure}

\begin{figure}
\resizebox{\hsize}{!}{\includegraphics{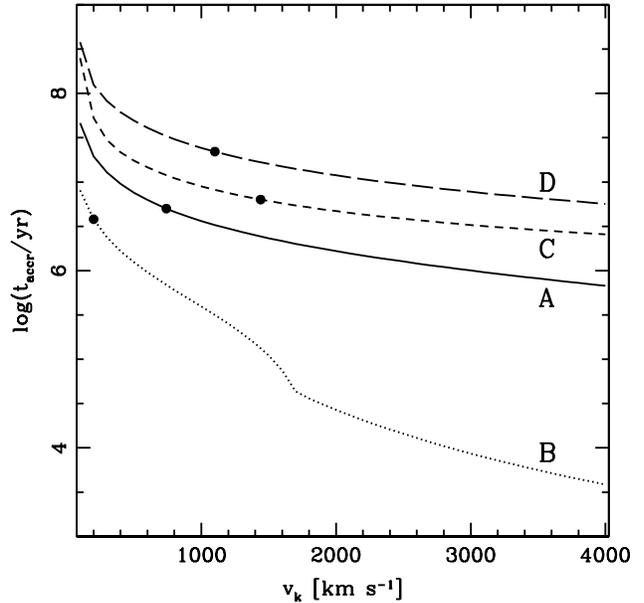}}
\caption[$t_{\rm accr}$ vs. $v_{\rm k}$ for all four galaxy models.]{$t_{\rm accr} = M_{\rm disk, ej}/\dot M_{\alpha}$ is plotted for $100 \leq v_{\rm k} \leq 4000$ km s$^{-1}$, for models A, B, C, \& D.  The solid dot on each curve denotes the escape velocity for that model.  The bump in curve B occurs when $r_{\rm ej} < r_{\rm Q1}$ and the disk profile changes.\label{fig:onlytaccr}}
\end{figure}

To quantify the duty cycles for off-center quasars, we plot both the simulation time $t_{\rm fin}$ (the time for the BH to settle back to the galactic center) and the calculated accretion time for the ejected BH disk, $t_{\rm accr}$, as functions of $v_{\rm k}$ (Fig.~\ref{fig:fid_taccr}).  The bold lines show the actual time for which the BH might be visible as an off-center quasar.  At low kick velocities, the BH settles back to the center before consuming its bound disk, and at high kick velocities, the BH accretes its disk and then continues to wander as a naked BH.  The effective time for off-center quasar activity, $\sim 10^{6.2} - 10^7$ yr, varies relatively little with $v_{\rm k}$.  This balance helps explain why the growth of the wandering BHs is also roughly constant with varying kick velocity (Figs.~\ref{fig:fid_dM},~\ref{fig:M6_dM},~\ref{fig:M9_dM},~\ref{fig:fgas_dM}).  However, Fig.~\ref{fig:onlytaccr} shows that $t_{\rm accr}$ continues to decrease, albeit slowly, for $v_{\rm k} > 800$ km s$^{-1}$.  Thus, we expect a lower $\Delta M/M_{\rm BH}$ for $v_{\rm k} \gg v_{\rm esc}$.

The simple picture shown in Fig.~\ref{fig:fid_taccr} provides a plausible explanation for the accretion behavior seen in our simulations based on off-center quasar activity alone, suggesting that accretion during disk crossings is relatively unimportant.  Also, for $v_{\rm k} \approx v_{\rm esc}$, $\Delta M/M_{\rm BH} \sim 10\%$ regardless of whether the BH escapes or returns for subsequent disk passages, further indicating that disk crossings contribute negligibly to the growth of recoiling BHs.  However, a more quantitative analysis demonstrates that this is not the complete picture.

\begin{figure}
\resizebox{\hsize}{!}{\includegraphics{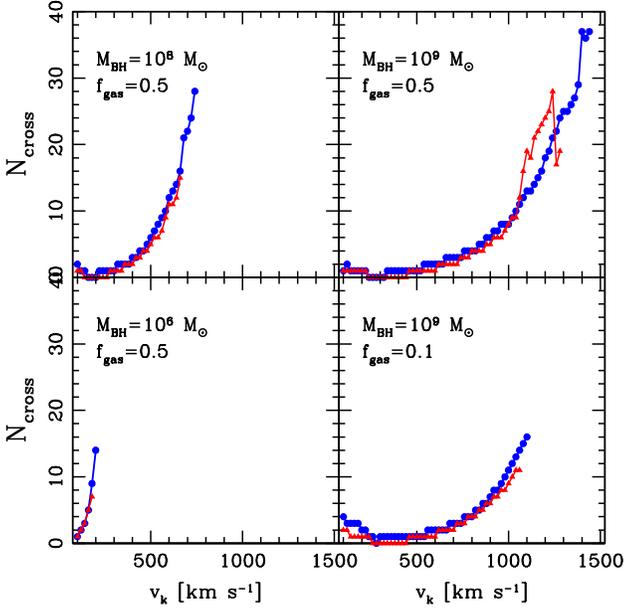}}
\caption[Number of disk crossings versus $v_{\rm k}$ for each model.]{ Number of disk crossings versus $v_{\rm k}$ for each model.  Blue with filled circles indicates kicks with $i = 45^{\rm o}$, and red with filled triangles indicates $i = 90^{\rm o}$.  {\it Upper left:} Model A.  {\it Lower Left:} Model B.  {\it Upper right:} Model C.  {\it Lower right:} Model D.  Only bound trajectories are shown, because the escaping BHs do not cross the disk. \label{fig:ncross}}
\end{figure}

As one would expect, the number of disk crossings ($N_{\rm cross}$) generally increases with increasing kick velocity (Fig.~\ref{fig:ncross}).  In each model, even Model B with the least massive BH, $N_{\rm cross} > 10$ for the highest kick velocities.  The kick inclination has little effect on the number of disk crossings.  However, as mentioned in \S~\ref{ssec:fidmod}, the amount of quasar activity during these crossings {\em is} inclination-dependent; virtually no accretion occurs when $i = 90^{\rm o}$, due to the small disk height in the inner regions.  Additionally, very low kick inclinations will produce few disk crossings; we therefore expect recoil kicks with moderate inclinations to be the most likely candidates for producing observable disk-crossing quasar activity.  We accordingly restrict our discussion of disk-crossing quasars to our runs with $i = 45^{\rm o}$.

\begin{figure}
\resizebox{\hsize}{!}{\includegraphics{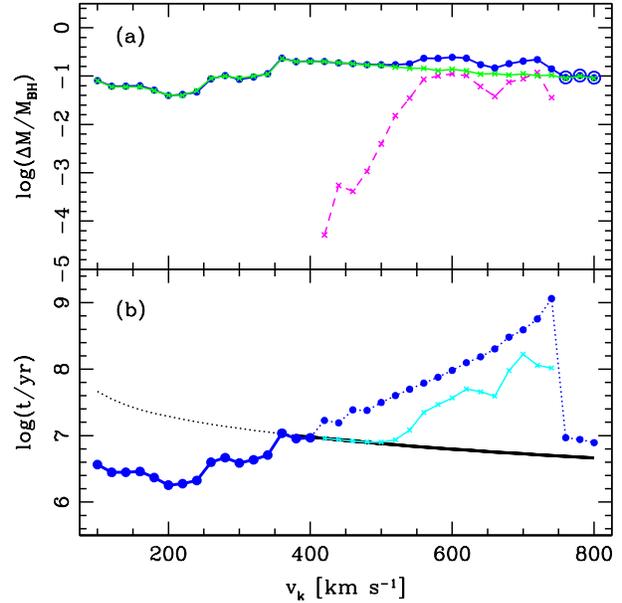}}
\caption[{\it (a)} $\Delta$M and {\it (b)} $t_{\rm accr}$ \& $t_{\rm fin}$ for each quasar type.]{{\it (a)} $\Delta M/M_{\rm BH}$ for Model A, $i = 45^{\rm o}$.  {\it Blue thick line with filled circles:} total mass accreted during simulation (same as in Fig.~\ref{fig:fid_dM}).  Open circles denote simulations in which BH escaped from the galaxy.  {\it Green thin line with crosses:} mass accreted from the ejected accretion disk, as an off-center quasar.  {\it Magenta dashed line with crosses:} mass accreted during disk crossings.  {\it (b)} Same plot as Fig.~\ref{fig:fid_taccr}, but with $i = 90^{\rm o}$ runs omitted and a ({\it cyan}) line with crosses added denoting the total quasar duty cycle for each simulation, i.e, $t_{\rm accr} + t_{\rm cross}$, where $t_{\rm cross}$ is the sum of duty cycles for all disk crossings in each run. \label{fig:allcross}}
\end{figure}

Figs.~\ref{fig:allcross}a \& b compare the contributions of both quasar types to the total quasar activity.  Fig.~\ref{fig:allcross}a shows the total mass accreted for the fiducial model, as well as the mass accreted via off-center quasars only and via disk crossings only.  We see that as expected, the off-center quasar phase dominates the overall growth, although at kick velocities $550\, {\rm km\, s^{-1}} \la v_{\rm k} < v_{\rm esc}$, the two contributions are comparable.  Note that in this same velocity range, the simulation time is generally greater than the expected timescale for a merger-induced starburst, $\sim 10^8$ yr.  Because accretion during disk crossings does not dominate the BH growth, however, a paucity of fuel in the post-starburst phase would reduce the total mass accretion by only a moderate amount.   Furthermore, if triggering of the starburst happens late in the merger process, the fuel supply may last longer.  

Fig.~\ref{fig:allcross}b is the same as Fig.~\ref{fig:fid_taccr} except that the $i = 90^{\rm o}$ runs are excluded, and an extra line is added to denote the total active quasar time for each simulation, $t_{\rm accr} + t_{\rm cross}$.  Interestingly, the total time spent in disk-crossing quasar phases is quite large at high velocities, up to $\sim$ 100 Myr in a few cases.  This result has two important caveats, however.  First, the accretion rate is clearly not Eddington-limited for the entirety of these long duty cycles, since the mass contribution from this type of accretion is at most comparable to that of off-center quasars.  Evidence for this sub-Eddington accretion can be seen in Fig.~\ref{fig:fid_i45v740}b.  {\rev Correspondingly, the luminosities of disk-crossing quasar events may often be relatively low and thus harder to detect than off-center quasar activity.  The second caveat is that} the longest duty cycles for disk crossings are those in which the BH has the lowest velocity, i.e., the final passages before the BH settles back to the galaxy center.  This is a tradeoff in observability: the disk-crossing quasars that would be most easily distinguished observationally from stationary quasars are those with the highest-velocity disk passages and, by definition, the shortest duty cycles. 

\begin{figure}
\resizebox{\hsize}{!}{\includegraphics{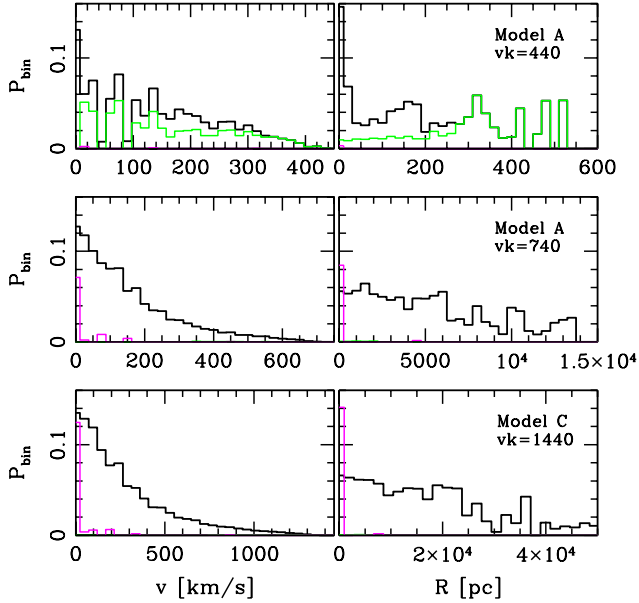}}
\caption[Velocity and spatial offset distributions for selected runs.]{Probability distributions of velocity ({\it left panels}) and spatial ({\it right panels}) offsets for selected runs.  In all cases, {\it black lines} denote the distribution for the entire simulation, {\it green lines} denote the distribution of time spent in off-center quasar phases only (i.e., $t < t_{\rm accr}$), and {\it magenta lines} denote the distribution of time spent in disk crossing phases only.  {\it Top panels:} Model A run with $i = 45^{\rm o}$ and $v_{\rm k} = 440$ km s$^{-1}$ (trajectory shown in Fig.~\ref{fig:fid_i45v440}).  {\it Middle panels:} Model A run with $i = 45^{\rm o}$ and $v_{\rm k} = 740$ km s$^{-1}$ (trajectory shown in Fig.~\ref{fig:fid_i45v740}).  {\it Bottom panels:} Model C run with $i = 45^{\rm o}$ and $v_{\rm k} = 1440$ km s$^{-1}$ (trajectory not shown).  Note that the probability distributions of off-center quasar phases in the higher-velocity runs are plotted, but are so low that they are almost invisible.\label{fig:rvdistns}}
\end{figure}

Fig.~\ref{fig:rvdistns} demonstrates this point more quantitatively; the distributions of velocity and spatial offsets are shown for three selected runs, with green and magenta lines denoting the time spent in off-center and disk-crossing quasar phases, respectively.  The top panels show the Model A run with $v_{\rm k} = 440$ km s$^{-1}$, in which the BH retains an accretion disk for about half of its wandering time.  The off-center quasar spends most of its lifetime at low velocities (near turnaround) and at spatial offsets of a few hundred parsec.  The BH crosses the disk only four times, however, so the time fraction it spends in disk-crossing quasar phases is very small.  The runs shown in the middle and bottom panels of Fig.~\ref{fig:rvdistns}, which have $v \la v_{\rm esc}$ for their respective models, {\rev have shorter off-center quasar duty cycles and longer wandering times, so the time fractions of the off-center quasar phases are very low -- virtually indiscernible.}  The time fraction spent in disk-crossing quasar phases is {\rev more} substantial, but because many disk crossings occur late in the simulation as the BH is settling back to the center, these are primarily constrained to the lowest velocity bin (and as expected, the smallest distance bin as well).  Therefore, while it is certainly possible in principle for GW recoil to produce a quasar with a measurable offset, it may prove challenging to observe one directly, especially one with a very high kick velocity.  

As mentioned in \S~\ref{sec:intro}, \citet{komoss08} {\rev have proposed that SDSSJ0927+2943 is} a recoiling quasar with a velocity offset of 2650 km s$^{-1}$.  We do not plot any recoil trajectories with kick velocities this high, because this is well above galactic escape velocities.  Were we to reproduce the distributions in Fig.~\ref{fig:rvdistns} for a run with $v_{\rm k} = 2650$ km s$^{-1}$, we would find a delta function at $v_{\rm k}$, because the BH would be ejected from the galaxy with little deceleration.  The quasar lifetime, however, would be short.  Taking the mass and luminosity values estimated by \citet{komoss08}, $M_{\rm BH} \sim 6\times 10^8$ M$_{\odot}$ and $L \sim 0.1 L_{\rm Edd}$, we find that the BH will only retain a gas disk of mass $\sim 2\% M_{\rm BH}$ and radius $\sim 0.37$ pc, which it will accrete within $\sim 10^7$ yr.  This is much shorter than the upper limit of $\sim 10^9$ yr that \citet{komoss08} estimate using $M_{\rm disk,ej} \approx M_{\rm BH}$.  If the {\rev conclusion} of \citet{komoss08} is accurate, then their discovery is extremely serendipitous, because it combines two low-probability events.  The first is the merger of two BHs with spins, mass ratio, and spin alignment configured in an unlikely manner to give a kick near the maximum possible GW recoil velocity, and the second is the observation of the recoiling BH during its {\rev relatively} short quasar phase.  {\rev This combined probability is difficult to calculate due to its dependence on unknown distributions of BH binary parameters, and it is unclear {\it a priori} whether this rare recoil event is in fact more or less likely to occur than a superposition of somewhat unusual quasars.  The possible role of gaseous outflows is also unclear.  Although outflows cannot easily explain all the unusual spectral lines, neither can the recoil scenario explain all of the narrow lines without outflows or similar phenomena being invoked at any rate.  Follow-up observations of this intriguing source would help to clarify the situation.}

Our finding that high-velocity GW recoil events are unlikely to be observed is consistent with the null result of \citet{bonshi07}, who conducted a search for velocity offsets $> 800$ km s$^{-1}$ in quasar spectra.  In Fig.~\ref{fig:onlytaccr}, we have plotted $t_{\rm accr}$ in all of our models, for the range $100 \leq v_{\rm k} \leq 4000$ km s$^{-1}$.  The decline in quasar duty cycle for high velocities can easily be seen, and at the highest velocities, all models have $t_{\rm accr} \la 10^7$ yr.  The duty cycle for an off-center quasar with $v_{\rm k} = 800$ km s$^{-1}$ is $6\times10^5 - 5\times10^6$ yr for our models in which $v_{\rm esc} < 800$ km s$^{-1}$ (Models A \& B), although the BH decelerates over this time.  The most massive BHs (Models C \& D) retain more mass after they are ejected and thus have longer duty cycles, but decelerate more quickly since they are on bound orbits at this $v_{\rm k}$.  Duty cycles are somewhat longer for these off-center quasars if the accretion luminosity is lower, although of course they then will be dimmer.  The duty cycles for disk-crossing quasars with $v \sim 800$ km s$^{-1}$ are shorter, however, $\sim 10^5$ yr.  In all cases, duty cycles are shorter still if one considers higher $v_{\rm k}$.  Clearly, the probability of observing a recoiling quasar is low, although future searches sensitive to smaller velocity offsets will increase these probabilities.  Finally, as a corollary to the arguments on the detectability of disk-crossing quasars, we note that \citet{kindeh05} and \citet{libesk06} have argued that some of the brightest ultraluminous X-ray sources (ULXs) might be explained as wandering intermediate-mass BHs ($M_{\rm BH} \sim 10^2 - 10^5 {\rm M_{\odot}}$) that were ejected from their host galaxies in previous merger events and that begin accreting again upon returning to a dense gaseous region.

There is another possibility for detecting recoil events in addition to observing quasar offsets.  \citet{lippai08}, {\rev \citet{shibon08}, and \citet{schkro08}} have recently pointed out that strong flares may occur following a GW recoil as the surrounding gas is disrupted and gas marginally bound to the BH falls back onto the galactic disk.  \citet{lippai08} argue that strong shocks are a natural consequence of perturbations to highly supersonic gas, as is the case with thin accretion disks.  If the shock propagates at $v_{\rm shock} \la v_{\rm k}$, it will produce a flare with an emission spectrum peaking at $k\, T_{\rm shock} = (3/16) \mu\, m_{\rm H}\, v_{\rm shock}^2 \approx 19\, {\rm eV}\, (v_{\rm shock}/100\, {\rm km}\, {\rm s}^{-1})^2$.  Such flares would be observable in soft X-rays for recoil kicks $\ga 200-300$ km s$^{-1}$; below these energies, most of the emission would be absorbed by interstellar hydrogen.  The highest kick velocities ($\ga 1000$ km s$^{-1}$) could produce hard X-ray flares, though these would be short-lived.  These photon energy ranges would change if the shock propagates at only a small fraction of the kick velocity, as \citet{lippai08} have estimated; in this case some flares might be observable at UV energies.  \citet{shibon08} estimate a typical flare timescale $t_{\rm flare} \sim 10^4$ yr and an observable flare rate of $\sim$ 1 in $10^4$ quasars.  {\rev \citet{schkro08} suggest that because the disk is optically thick, recoil flares may produce long-lived infrared afterglows, of which up to $\sim 10^5$ might be visible today.}  These numbers are uncertain, however, and detailed hydrodynamical simulations of recoiling disks will be needed to more accurately determine the frequency and duration of these flares, as well as whether their spectra and luminosities can be distinguished easily from concurrent AGN activity.  The important point to note, however, is that a recoiling BH making numerous passages through the disk may create a succession of flares, rather than just an initial flare during the recoil event.  This could substantially increase the rate at which recoil flares occur.

\subsection{Sensitivity of results to $q$ parameter}
\label{ssec:q_sensitivity}

\begin{figure}
\resizebox{\hsize}{!}{\includegraphics{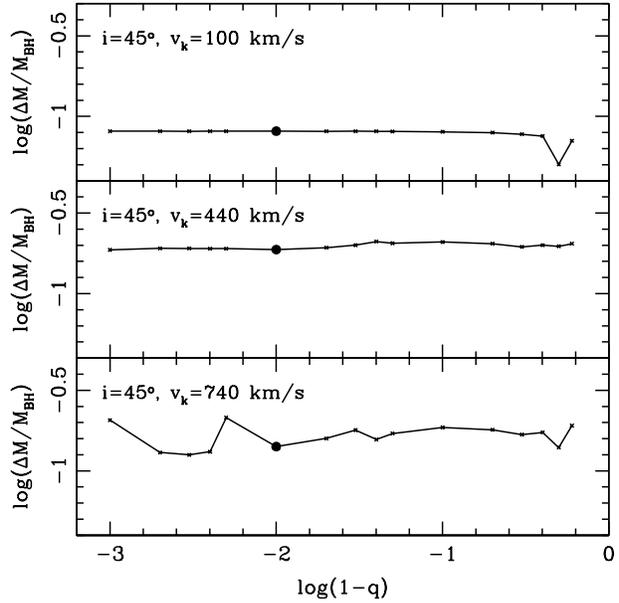}}
\caption[$\Delta M/M_{\rm BH}$ for various values of $q$.]{$\Delta M/M_{\rm BH}$ as a function of the ellipticity parameter $q$, using Model A and $i=45^{\rm o}$.  The low ($v_{\rm k} = 100$ km s$^{-1}$, {\it top panel}), moderate ($v_{\rm k} = 440$ km s$^{-1}$, {\it middle panel}), and high-velocity ($v_{\rm k} = 740$ km s$^{-1}$, {\it bottom panel}) kicks shown in \S~\ref{ssec:fidmod} are chosen again here to demonstrate variation with $q$.  For clarity, the value log($1-q$) is plotted on the x-axis.  The solid dot denotes our chosen fiducial value $q=0.99$.\label{fig:deltaM_q}}
\end{figure}

As mentioned above, we have accounted for the ellipticity of the gas disk potential $\Phi$(R) by introducing a small perturbation, $q$, in the radial coordinate such that $\tilde R = \sqrt{ r^2 + (z/q)^2 }$ and $\Phi = \Phi(\tilde R)$.  We use $q = 0.99$ as a fiducial value for our runs, and here we investigate the sensitivity of our results to this choice of $q$.  {\em A priori}, one can expect the largest perturbation to the BH orbit to occur near turnaround, where the velocity is low and the BH is most sensitive to acceleration from the (perturbed) gravitational potential.   Therefore, the largest cumulative perturbation should occur for orbits with the most turnarounds, i.e. those with large $v_{\rm k}$.  Orbits with large $v_{\rm k}$ also have longer turnaround times, since turnaround can be defined as a fixed fraction of the orbital period, which increases for higher kick velocities.  Fig.~\ref{fig:deltaM_q} shows the total mass accretion, $\Delta M/M_{\rm BH}$, as a function of $q$ for low, moderate, and high values of $v_{\rm k}$.  Indeed, we see that  very little variation with $q$ occurs except in the highest velocity case, $v_{\rm k} \la v_{\rm esc}$.  We can thus conclude that the value of $q$ does not matter for simulations with low-to-moderate $v_{\rm k}$.  For the largest kicks, we can assume that choosing $q$ fairly close to unity is appropriate, since turnarounds will occur mostly at large radii where the degree of ellipticity of equipotential contours should be small.  Furthermore, BH trajectories for $v_{\rm k} \la v_{\rm esc}$ are already the most susceptible to other inaccuracies due to physical effects that were not included here.  For example, at the largest radii probed by orbits bound to the baryonic galactic component (generally tens of kpc), BHs are likely to encounter perturbations due to the triaxiality of the DM halo.  The host galaxy is also likely to evolve and undergo star formation during the the long ($\ga$ Gyr) wandering time of the BH in these cases.  

\section{Discussion}
\label{sec:disc}

We have considered SMBH accretion without requiring that the SMBH remains stationary at the center of its host galaxy.  Instead, the SMBH was assumed to be ``kicked" from the center at 100s - 1000s of km s$^{-1}$, which can occur due to GW recoil following a BH merger.  To calculate the trajectories of recoiling SMBHs in a galactic potential, we have developed an analytical model that includes a stellar bulge and a multi-component gaseous disk.  The SMBH mass and gas fraction are free parameters that we varied in four different cases (Models A-D).  For bound trajectories, we examined the dynamics and the wandering timescales of the SMBHs, as well as the effect of this wandering on SMBH growth.

Surprisingly, the kick velocity, $v_{\rm k}$, has little effect on the amount of mass accreted onto the SMBH while it wanders through the galaxy.  For all cases, the fractional SMBH growth, $\Delta M/M_{\rm BH}$, is roughly 10\% over the time from the recoil kick to when the BH settles back to the galactic center.  This is primarily the combined result of two different effects:  the total wandering time is short for low $v_{\rm k}$, where the BH quickly settles back to the center, and the accretion time is short for high $v_{\rm k}$, where the accretion disk bound to the BH is small.  Due to these competing factors, the timescale for off-center accretion, and hence the amount of growth in this time, remains almost constant for all $v_{\rm k}$.  Additionally, bursts of accretion during subsequent disk crossings contribute little to the total BH mass.  The consistency of $\Delta M/M_{\rm BH}$ implies that GW recoil is an effective means of self-regulation for SMBH growth.  If the SMBH remained stationary at the galactic center, substantial growth could occur.  This growth could be regulated via more conventional processes such as energy or momentum feedback, but these depend on the details of the gas physics and may vary greatly in different galactic environments.  In contrast, GW recoil depends only on the parameters of the merging SMBH progenitors.  It is also interesting to note that because other feedback processes will be halted when the recoiling SMBH has depleted its reservoir of bound gas, these two methods of growth regulation are complementary.  The possibility that both wandering and stationary SMBHs self-regulate their masses indicates how recoiling BHs could be consistent with the observed $M_{\rm BH} - \sigma_*$ relation, even if large recoils $\la v_{\rm esc}$ are not rare occurrences.  The latter conclusion assumes, of course, that the escape fraction of recoiling SMBHs is still small; otherwise we would expect to see more scatter in the $M_{\rm BH}-\sigma_*$ relation {\rev \citep{libesk06, volont07}} or galaxies without a central SMBH in the local universe, which are not observed \citep[e.g.,][]{korric95, richst98, ferfor05}.

Even small recoil kicks can have a significant impact on accretion and feedback processes.  For the lowest kick we consider, $v_{\rm k} = 100$ km s$^{-1}$, the BH wanders in the galaxy core for a few Myr before settling back to the center and carries along a large accretion disk.  Thus, feedback from moving sources will be distributed over much larger volumes than the emitting regions of stationary AGN.  For example, in our fiducial model, a BH kicked at 100 km s$^{-1}$ reaches $\sim$ 30 pc before turning around.  Assuming most of the AGN emission originates from a region $\sim 10^2 - 10^3$ $r_{\rm S}$ across, the initial volume in which energy is released by the BH increases by about 8 orders of magnitude in this case.  Most of the energy will be deposited at the largest radii, since turnaround takes the largest fraction of the orbital time.  This implies an even greater disparity between the impact of moving and stationary BHs on their surrounding environments.  

The importance of GW recoil in the evolution of SMBHs clearly depends on how often significant recoil kicks occur.  This is difficult to estimate, because the recoil kick distribution depends on the spin and mass ratio distributions of the progenitor BH binaries, which are not well known.  {\rev \citet{schbuo07}, \citet{campan07a}, \& \citet{baker08} have calculated kick distributions; \citet{schbuo07} use an approximate method, and the latter two derive empirical formulae from numerical relativity simulations.  The differences in the results of the latter two stem from different assumptions of how the kick velocity scales with BH progenitor mass ratio.  Assuming BH spins of $a=0.9$, mass ratios in the range $1/10 \leq q \leq 1$, and random spin orientations, the three groups respectively estimate that ($12, 36, \&\, 23\%$) of recoiling BHs will have $v_{\rm k} > 500$ km s$^{-1}$ and ($3, 13, \&\, 9\%) > 1000$ km s$^{-1}$.  Note that these fractions are not insignificant, although they might be much smaller if BH spins are on average lower or are preferentially aligned \citep[cf.][]{bogdan07}.}  To constrain the actual recoil kick distribution, better observational constraints on merging BHs or observations of individual recoiling BHs will be required.  

Our results indicate that the latter method for constraining GW recoil, i.e. direct observations of recoiling BHs, will be challenging, because several factors make recoiling BHs difficult to observe.  First, the BH must be actively accreting to appear as an observable source.  The recoil could be observed either via spatial offsets from the galactic nucleus or via velocity offsets in the quasar spectrum.  In both cases, larger offsets resulting from large recoil kicks are more easily observable, but large kicks imply a small disk bound to the BH and thus a short off-center quasar duty cycle.  After it exhausts its supply of bound gas, the moving BH could again become an active source on subsequent passages through the galactic gas disk, an effect which could be observed as a velocity offset in its spectrum.  {\rev This could also be a source of periodic quasar or disk flare activity.}  However, the greater the velocity offset, the shorter the duty cycle for these ``disk crossing" quasars, since they pass rapidly through the relatively thin galactic disk.  It is worth emphasizing that because our results indicate that GW recoil may be difficult to observe directly and that the growth of wandering BHs may be consistent with observed galaxy-BH relations, neither of these observational methods may be able to rule out a significant population of recoiling BHs.

If, indeed, GW recoil is a relatively common phenomenon in SMBH mergers, this represents a fundamental change in our understanding of SMBH-galaxy coevolution, because we cannot gain a complete picture by considering only stationary SMBHs.  In reality, GW recoil and AGN feedback are likely to work together to produce the observed SMBH-galaxy relations.

\section*{acknowledgements}
{\rev We thank Bence Kocsis, Gregory Shields, and Ryan O'Leary for useful comments.}  This work was supported in part by Harvard University funds.  

\bibliography{refs}

\end{document}